\documentstyle[emulateapj]{article}

\def\lae{\mathrel{<\kern-1.0em\lower0.9ex\hbox{$\sim$}}}
\def\gae{\mathrel{>\kern-1.0em\lower0.9ex\hbox{$\sim$}}}

\def\tt{\tablenotetext}

\slugcomment{Accepted for publication in the Astronomical Journal, October 1999}
 
\lefthead{C\^ot\'e et al.}
\righthead{Andromeda II}
 
\begin{document}
 
\title{Abundances of Red Giants in the Andromeda II Dwarf Spheroidal Galaxy}
 
\author{Patrick C\^ot\'e\altaffilmark{1}}
\affil{California Institute of Technology, Mail Stop 105-24, Pasadena, CA 91125\\
    {\rm pc@astro.caltech.edu}} 
\centerline{~}

\author{J.B. Oke}
\affil{California Institute of Technology, Mail Stop 105-24, Pasadena, CA 91125, and\\
    Dominion Astrophysical Observatory, Herzberg Institute of Astrophysics,
    National Research Council of Canada, 5071 W. Saanich Road, Victoria, BC, V8X 4M6, Canada\\
    {\rm Bev.Oke@hia.nrc.ca}}

\and

\author{Judith G. Cohen}
\affil{California Institute of Technology, Mail Stop 105-24, Pasadena, CA 91125\\
{\rm jlc@astro.caltech.edu}}

\altaffiltext{1}{Sherman M. Fairchild Fellow}

 
 
\begin{abstract}
We have obtained spectra for 50 candidate red giants in Andromeda II, a dwarf spheroidal
companion of M31, using the Low Resolution Imaging Spectrometer on the Keck II telescope.
After eliminating background galaxies and Galactic foreground 
stars, we are left with a sample of 42 red giants for which membership in Andromeda II
can be established unambiguously from radial velocities.
Line indices measured on the Lick/IDS system 
are combined with $VI$ photometry obtained with the Keck II and Palomar 5m telescopes 
to investigate the age and metallicity distribution of these stars.
Based on a comparison of the measured line indices to those
of Lick/IDS standard stars in globular and open clusters, we derive 
a mean metallicity of ${\overline{\rm [Fe/H]}} = -1.47\pm0.19$ dex.
This confirms earlier conclusions based on Thuan-Gunn $gr$ photometry that 
Andromeda II obeys the familiar relation between mean stellar metallicity and galaxy 
luminosity. There is also evidence for a dispersion in metallicity 
of $\sigma$([Fe/H]) = $0.35\pm0.10$ dex based on the scatter in the measured Mg$b$
line indices and the observed width of the galaxy's giant branch. We note
that while existing observations of Local Group dwarf galaxies indicate that
their mean metallicity depends rather sensitively on total luminosity, the 
internal {\sl spread} in metallicity appears to be relatively independent of
galaxy luminosity.

Our spectroscopic sample contains one carbon star. We measure $M_I \simeq -3.8$ 
for this star, which places it below the tip of the red giant branch and suggests 
a common origin with the CH stars found in the Galactic halo. Although this carbon 
star alone does not provide evidence of an intermediate-age component in Andromeda 
II, two other stars in our spectroscopic sample have $M_I \simeq -4.7$ and $-4.5$.
Membership in Andromeda II is unambiguous in both cases, indicating
that these stars fall along an extended asymptotic giant branch and pointing to the
presence of a modest intermediate-age population in this galaxy.
\end{abstract}
 
 
 
 
%

\section{Introduction}

Andromeda II (And II) is a faint dwarf spheroidal (dSph) galaxy located approximately
10$^{\circ}$ from the center of M31. It was discovered, along with 
two other dSph companions of M31, by van den Bergh (1972) who visually
searched an area of $\sim$ 700 deg$^2$ using plates taken with the Palomar Schmidt 
telescope. Recently, three new M31 dSph galaxies have been discovered,
bringing the total number of dSph galaxies associated with M31 to six (Armandroff, Davies \&
Jacoby 1998; Armandroff, Davies \& Jacoby 1999; Karachentsev \& Karachenseva 1999).
Although these systems have been studied to differing extents,
existing observations suggest that they bear a remarkable similarity to the
dSph galaxies which belong to the Milky Way ($e.g.$, Da Costa et al. 1996; Armandroff
et al. 1998; Armandroff et al. 1999; Hopp et al. 1999; Grebel \& Guhathakurta 1999;
Caldwell 1999).

Several curious properties of the Galactic dSphs motivate the 
study of dSphs associated with other galaxies. First, their large central velocity
dispersions indicate that they contain significant dark matter components (Aaronson 1983;
Faber \& Lin 1983; Mateo 1998).  Second, color-magnitude diagram (CMD) studies reveal
detailed and extraordinarily varied star formation histories
(Da Costa 1992; Smecker-Hane 1994, Stetson, Hesser \& Smecker-Hane 1998). 
Last, in many cases, the observed metallicity distribution functions indicate
surprisingly wide ranges in metallicity and thus point to complex
chemical enrichment histories (Canterna 1975; Zinn 1978; Shetrone, Bolte \& Stetson 1998).

The presence of intermediate-age stars in these galaxies 
complicates the derivation of metallicity distribution functions based on
broad-band photometry alone. For most of the Galactic dSphs, spectroscopic 
metallicity determinations for individual red giants are now available
($e.g.$, Da Costa et al. 1991; Suntzeff et al. 1993; Ibata et al. 1997).
However, existing constraints on the metallicities of the dSphs associated with M31 come
entirely from broad-band photometry (Armandroff et al. 1993; Da Costa et al. 1996; 
Hopp 1999; Grebel \& Guhathakurta 1999). In
the case of And II, the sole metallicity determination
comes from Thuan-Gunn $gr$ photometry obtained with the four-shooter CCD camera
on the Palomar 5m telescope
(K\"onig et al. 1993; hereafter KNMF). Based on the location and width of the 
red giant branch (RGB), these authors reported a mean metallicity of 
${\overline{\rm{[Fe/H]}}} = -1.59^{+0.44}_{-0.12}$ dex and
an internal dispersion of ${\sigma}$([Fe/H]) $\sim$ 0.43 dex.
However, the possible existence of intermediate-age stars in this galaxy
(Aaronson et al. 1985) raises concerns that spreads in both metallicity
and age may be contributing to the observed width of the giant branch.
Clearly, spectroscopic information and/or ultra-deep {\sl HST} imaging 
(such as that presented by Da Costa et al. 1996 for And I) is required to 
break the well known ``age-metallicity degeneracy" in this and other M31 dSphs.
In this paper, we investigate the chemical abundances of individual red giants in
And II using intermediate-resolution ($R \simeq 6$ \AA ) spectra obtained
with the Keck II telescope; this is the first such study
for any dSph galaxy beyond the Milky Way.

\section{Observations and Reductions}

\subsection{Photometry and Astrometry}

Candidate red giants in And II were selected from a single 600-second
$V$ image of And II obtained on 17 August 1996 using COSMIC ($i.e.,$ 
Carnegie Observatories Spectroscopic Multi-slit and Imaging Camera; Kells
et al. 1998) on the Palomar 5m telescope. 
COSMIC was used in direct imaging mode, giving a total field of view of 
9\farcm7$\times$9\farcm7. The FWHM of isolated stellar objects in this image 
was measured to be $\sim$ 1\farcs2.  The stand-alone version of DAOPHOT II 
(Stetson 1987; Stetson 1993) was used to measure instrumental magnitudes
for 2089 unresolved objects in this field, and stellar objects within the upper
$\simeq$ one magnitude of the RGB were randomly selected for spectroscopic observation 
with the Keck II telescope. In a few cases, somewhat brighter objects were added 
to fill the LRIS slit mask.
Absolute positions for these candidate red giants were calculated using
the positions of 32 bright stars taken from the USNO-A2.0 Catalog (Monet et al. 1996).

On 7 October 1996, $V$ and $I$ images of And II were taken with 
the Low Resolution Imaging Spectrometer (LRIS; Oke et al. 1995) on the 
Keck II telescope. Exposure times were 600 seconds in $V$ and 300 seconds in $I$. 
The FWHM of unresolved objects in these images was measured to be 0\farcs88 and 
0\farcs76, respectively. The images were bias-subtracted,
trimmed and flat-fielded using median sky flats obtained
during twilight. Profile-fitting photometry was performed
using the stand-alone version of DAOPHOT II. Unfortunately, no photometric standard 
stars were observed on this night, so it was not possible to calibrate the
LRIS photometry directly.
Instead, the Keck photometry was calibrated by re-observing And II and four
Landolt (1992) standard fields using the Palomar 5m telescope and COSMIC on 
the night of 24 July 1998.  Since COSMIC is not equipped with a Cousins $I$ 
filter, a Thuan-Gunn $i$ filter was used as a substitute. 
Based on the residuals between our calculated and observed
$V$ magnitudes and $(V-I)$ colors of the Landolt standards, we find standard
deviations of 0.04 mag ($V$) and 0.07 mag ($V-I$).

\subsection{Spectroscopy}

Two LRIS masks were designed containing a total of 52 slits: 28 on the 
first mask and 24 on the second, with two stars (\#11 and 20)
included on both masks. On 7 October 1996, we obtained a pair of
3600-second exposures for both masks using a 600 l/mm grating blazed at 5000 \AA . 
This configuration produced a dispersion of 1.28 \AA\ pixel$^{-1}$ and, when 
combined with our slit width of 1\farcs0, a resolution
of approximately 6 \AA\ over the range 3950 to 6000 \AA . 
The seeing during the observations was measured to be
FWHM $\sim$ 0\farcs8.
Following each exposure, we obtained a comparison spectrum of
Hg, Kr and Ar lamps which was subsequently used to derive the 
dispersion solution of the program spectra.
Crude radial velocities were measured by cross-correlating the spectrum of 
each program object against those of red giants in the globular clusters 
M13, M92 and M71. The radial velocities have a typical uncertainty of 
$\sigma_{v_r}$ $\sim$ 40 km s$^{-1}$ which, given the heliocentric radial velocity
velocity of ${\overline{v}}_r = -188\pm3$ km s$^{-1}$ for And II
(C\^ot\'e et al. 1999), is sufficient for establishing membership. For 
objects \#3, 31 and 46, no radial velocity could be measured;
these stars are omitted from the determination of the metallicity of And II. 

Table 1 gives the identification number of each candidate red giant observed
spectroscopically with LRIS, its right ascension and declination,
$V$ magnitude and $(V-I)$ color. For those objects which are located in both
our LRIS field and the four-shooter field of KNMF, Table 1 includes the $g$
magnitude, $(g-r)$ color, and star identification number from KNMF. The
final column of Table 1 indicates whether the object is a member of And II
based on its measured radial velocity (see below). A finding chart for
all objects listed in this table is presented in Figure 1.

Figure 2 shows extracted, wavelength-calibrated and co-added spectra for
six of the And II member giants. The strong C$_2$ bands in the spectrum of star \#3
immediately identifies it as a carbon star. Aaraonson et al. (1985) reported the 
presence of carbon stars in And II, and suggested on this basis that
And II contains a modest fraction of intermediate-age stars. We return
to the issue of intermediate-age stars in \S 3.2.

\section{Results}

\subsection{Adopted Distance and Reddening}

The primary goal of this study is the determination of
spectroscopic metallicities for individual And II red giants.
To do so, we utilize color and reddening information to determine
the temperature differences between the program stars and
the Lick/IDS standard stars whose line indices define the
metallicity scale (see \S 4). Figure 3 shows the $I,(V-I)$ CMD
for all unresolved objects within 1$^{\prime}$ of the galaxy's center.
The core radius of And II is $r_c$ = 1\farcm89$^{+0.47}_{-0.37}$
(Caldwell et al. 1992; C\^ot\'e et al. 1999), so this selection ensures
that the vast majority of the objects plotted in Figure 3 are bonafide
And II members. Also shown are globular cluster fiducial giant branches
from Da Costa \& Armandroff (1990). In the left panel of Figure 3, these have been
shifted by $(m-M)_I = 24.22$ and
$E(V-I) = 0.08$. These values are appropriate for a Galactic foreground
reddening of $E(B-V) = 0.062$ (Schlegel, Finkbeiner \& Davis 1998),
$E(V-I) = 1.36E(B-V)$ (Taylor 1986; Fahlman et al. 1989),
$A_I = 1.86E(B-V)$ and a true distance modulus of $(m-M)_0 = 24.1$.
In the right panel, we show the results of adopting $(m-M)_0 = 23.9$,
$A_I = 0.15$ and $E(B-V) = 0.08$ (the reddening assumed by KNMF).
In what follows, we adopt the reddening deduced from the DIRBE maps
of Schlegel, Finkbeiner \& Davis (1998)
and $(m-M)_0 = 24.1\pm0.3$, where the rather large uncertainty reflects
both the uncertainty in the true reddening toward And II and potential
systematic errors in our photometric calibration. 

The corresponding distance of $D_{\rm And II} = 660\pm90$ kpc places
And II $\sim$ 85$\pm$90 kpc in front of M31, for an adopted distance of
$D_{\rm M31} = 745$ kpc (see \S 1 of Holland 1998). Based the
apparent magnitude of the tip of the RGB in And II, KNMF concluded that it
is located $\sim$ 120 kpc closer than M31, for an assumed M31 distance of
$D_{\rm M31} = 700$ kpc. We consider this agreement acceptable given the 
uncertainties involved in transforming our Thuan-Gunn $i$ photometry
to the Cousins system.

\subsection{Intermediate Age Stars?}

Aaronson et al. (1985) presented low-S/N spectra for four luminous giants in And II
obtained with the Palomar 5m telescope. Their sample includes
one unambigous carbon star (A211), one possible carbon star (A10) and one
M giant (A209). Based on $JHK$ photometry for these stars and an {\sl assumed}
distance modulus of $(m-M)_0$ = 24.3 for And II, Aaronson et al. (1985) derived 
bolometric magnitudes of $-4.45 \lae M_{\rm bol} \lae -4.1$ for these stars and
concluded that And II contains at least some intermediate-age stars.

Since CH carbon stars are found in the Galactic halo, as well as in at least two 
globular clusters ($\omega$ Cen and M14; see Harding 1962; Dickens 1972;  C\^ot\'e et al. 
1997), the mere existence of carbon stars in And II does not provide 
unambiguous evidence of an intermediate-age population (although the {\sl absence} of
such a population in And II might perhaps be surprising given the emerging evidence
for intermediate-age populations in the vast majority of Local Group dSphs; see 
Mateo 1998; Grebel 1999). On the other hand, carbon stars with $M_I \lae -4.0$ 
(Da Costa \& Armandroff 1990) have luminosities which exceed those of the
brightest red giants found in metal-poor globular clusters, so it is safe to assume that 
they are, in fact, intermediate-age objects which have undergone carbon dredge-up
during the ascent of the asymptotic giant branch (AGB), and not the end-products 
of mass-transfer evolution among compact binaries (McClure 1997).

None of the four stars observed by Aaronson et al. (1985) were included in 
our spectroscopic sample, although we include these objects in the finding chart 
given in Figure 1. Curiously, the two certain carbon stars in And II ($i.e.$, our \#3 and 
Aaronson \#211) are separated by only 13$^{\prime\prime}$. The location 
of the four stars studied by Aaronson et al. (1985) in the CMD 
is shown by the filled squares in Figure 4. We find
$-5.0 \lae M_I \lae -4.2$ for these four stars, confirming the principal conclusion of 
Aaronson et al. (1985) that these objects belong to an extended AGB population.
On the other hand, the newly discovered carbon star 
has $M_I \simeq -3.8$ and consequently does {\sl not} lie above tip of the RGB. 
We note that two other member
stars (objects \#14 and \#30 which have $M_I \simeq -4.7$  and $-4.5$, respectively) are 
also located above the RGB tip in Figure 4 and are probably, like the 
four stars identified by Aaronson et al. (1985), members of an 
intermediate-age component in And II.

\section{Mean Metallicity and Abundance Spread}

\subsection{The Color-Magnitude Diagram}

Based on the CMDs of And II presented in Figures 3 and 4, we 
estimate a mean metallicity of ${\overline{\rm [Fe/H]}} \sim -1.35\pm0.3$ dex.
This is marginally higher than the 
value of ${\overline{\rm [Fe/H]}} \sim -1.59\pm^{+0.44}_{-0.12}$ dex reported by 
KNMF, although still consistent within the rather large uncertainties.
In addition, we find the width of the RGB to be significantly
larger than that expected purely on the basis of the photometric errors.
The left panel of Figure 5 shows the $I,(V-I)$ color-magnitude diagram
for 134 unresolved objects having $21 \le I \le 22$ and lying within one
arcminute of the galaxy's center. The dotted lines show the fiducial globular
cluster red giant branches shown in Figures 3 and 4; the solid line
indicates our adopted ridge line for And II. The right panel of
Figure 5 shows the histogram of $(V-I)$ color residuals about this ridge line. The
best-fit Gaussian, which has a dispersion of ${\sigma}(V-I)$ = 0.11$\pm$0.03 mag, has
been overlaid for comparison. Since the mean internal photometric uncertainty
for objects within this magnitude range is $\sigma(V-I)$ = 0.05 mag, we conclude
that the intrinsic dispersion in color is roughly $\sigma(V-I)$ $\simeq$ 0.10 mag.
At the median magnitude of $I = 21.62$ mag for these stars, the gradient in $(V-I)$ 
color as a function of metallicity is approximately 0.21$\pm$0.04 mag dex$^{-1}$. Thus, 
if the broadening of the RGB is ascribed entirely to an internal spread in
metallicity, inferred dispersion is $\sigma({\rm{[Fe/H]}}) \sim 0.46\pm0.17$ dex.
Note that this estimate is unlikely to be contaminated by the presence of
old, metal-rich AGB stars since for metallicities similar to that of Anda II such stars
appear in significant numbers for $M_I \gae -1.5$ (see Figure 5 of Ferraro et al. 1997)
whereas our adopted limits correspond to absolute magnitudes of 
$-3.2 \lae M_I \lae -2.2$.
Nevertheless, the presence of at least some intermediate-age stars in And II suggests that 
is probably wise to interpret the above estimate an as upper limit on the true metallicity 
dispersion (c.f. Smecker-Hane et al. 1994 who demonstrates that, at least in the case of the 
Carina dSph, the RGB is relatively narrow despite Carina's multiple star-formation 
episodes).

\subsection{Line Index Measurements}

During Keck/LRIS observing runs in April 1996, October 1996 and April 1999, we 
collected high-S/N, longslit spectra for 12 Lick/IDS standard stars
in M13, M92, M71 and M67.
For M13, long-slit spectra were also measured
for eight additional red giants having published $(V-K)_0$ colors (Cohen, Persson \& Frogel 1978).
In all cases, the identical grating and central wavelengths were used as for the 
And II observations, although slits of differing widths were employed for
the three runs: $i.e.$, 0\farcs7, 1\farcs5 and 1\farcs0 for the October 1996,
April 1996 and October 1999 observations, respectively.
These slits
correspond to spectral resolutions in the range $4 \lae R \lae 9$ \AA\ which are
roughly comparable to the resolution of $R = 8$  \AA\ used to define the Lick/IDS 
indices (Worthey et al. 1994).
All 18 stars are listed in Table 2
which gives the identification number of each star, its absolute magnitude, $(V-K)_0$
and $(V-I)_0$ colors. 
These quantities have been taken directly from 
Cohen et al. (1978), Gorgas et al. (1993) and Worthey et al. (1994), 
with the exception of $(V-I)_0$ which has been estimated for each star
using equation (10) of von Braun et al. (1998).  The mean metallicity of the host
cluster, taken from the catalog of Harris (1996), is indicated in the final column.
Using the index definitions given in
Worthey et al. (1994), we have measured absorption line indices for each of
the stars listed in Table 2. Figure 6 shows our measured line indices for the
12 Lick/IDS standards plotted against the values given in Worthey et al.  (1994). 
The NaD index was also measured using our program spectra, although it is
not considered here since its use is complicated by the existence of
interstellar sodium, whose column density is
also known to vary significantly over small scales (Cohen 1979; Gebhardt et al. 1994).
The two datasets are in good agreement, with the possible exception of
a 0.015 mag offset for the Mg$_1$ and Mg$_2$ indices (in the
sense that the Keck spectra yield slighly higher values). The sample of objects
is small, however, and we conclude from Figure 6 that our measured indices are in
satisfactory agreement with the published values. 

Table 3 lists measured line indices of 41 red giants belonging to
And II. For the Mg$_1$, Mg$_2$, Mg$b$ and Fe5406 indices, we have imposed a 
minimum S/N in the local continuum of 15 per pixel; for G4300, which is
the bluest index, this threshold was relaxed to S/N = 10 per pixel.
The uncertainties in the measured indices have been estimated by adding, in
quadrature, the uncertainty due to Poisson noise and with that due to
slit alignment errors. The latter is the dominant uncertainty for our
measured line indices since the photon noise in our LRIS spectra never
exceeds $\sim$ 10\% pixel$^{-1}$ at
5000 \AA . We estimate the uncertainty due to slit alignment errors by shifting
each spectrum by
$\pm \delta$, where $\delta$ is half the width of
the slit in wavelength units ($e.g.$, $\delta \sim 3.0$ \AA ), and re-measuring
the value of each line index; the uncertainty is then taken to be
half the range spanned by the measurements.

In general, the measured index for any star will be a function of not just
metallicity, but also effective temperature and surface gravity. Figure 7
shows the variation in surface gravity as a function of absolute magnitude
expected for old, metal-poor, red giants based on the models of Bergbusch 
\& VandenBerg (1992). All of the stars for which we have measured 
line indices lie within the upper $\sim$ one magnitude of the And II
giant branch, suggesting that these stars have surface gravities in the
range $0.4 \lae {\rm log}~g \lae 1.2$, with 
${\rm log}~g \simeq 0.8$ being typical.

Figures 8-12 show the measured values of the G4300, Mg$_1$, Mg$_2$, Mg$b$ and Fe5406
indices plotted against $(V-I)_0$ for the confirmed And II members.
Also shown are line indices for Lick/IDS standard stars in a sample
of nine globular and open clusters which span a wide range in metallicity,
supplemented by new measurements for the eight red giants in
M13 listed in Table 2. For the standard
stars, we have transformed the published $(V-K)_0$ colors
to $(V-I)_0$ using equation (10) of von Braun et al. (1998).
In each figure, we show the corresponding Lick/IDS fitting functions 
from Gorgas et al. (1993) and/or Worthey et al. 
(1994).\altaffilmark{2}\altaffiltext{2}{These fitting functions
are based primarily on the globular cluster abundance scale of Kraft (1979),
whereas the metallicity scale of Harris (1996) was used to calibrate the $VI$ 
giant branch technique employed in \S 4.1. We note, however, that for the
six globular clusters used in the calibration by Lick fitting functions, 
the two scales show a mean difference of only ${\Delta}$[Fe/H] = 0.03$\pm$0.10 
dex.} The dashed lines indicate the expected relations for ${\rm log}~g = 0.8$ and 
five different choices of metallicity: Fe/H] = $-$2.5, $-$2.0, $-$1.5, $-$1.0 
and $-$0.5 dex.  As a demonstratation
of the gravity sensitivity of each index, the solid lines indicate
the relations for the three cases of ${\rm log}~g$ = 0.4, 0.8 and 1.2,
and [Fe/H] = $-1.67\pm$0.16 dex. As explained below, this metallicity
produces the best simultaneous match between the Lick/IDS fitting functions and 
the combined set of line indices measured for the And II program stars.

For each line index, we have computed the $\chi^2$ statistic for the appropriate 
Lick/IDS fitting function over the range $-2.5 \le$ [Fe/H] $\le 0.0$ dex in 
increments of $\Delta$[Fe/H] = 0.025 dex. 
The value of [Fe/H] which produced the lowest
$\chi^2$ was then adopted as the best-fit mean metallicity.
The small number (3-5) of very red 
stars having effective temperatures outside the range in which the fitting 
functions apply (as recommended by Gorgas et al. 1993 and Worthey et al. 1994) 
have been omitted from the $\chi^2$ calculation.
All measurements have been assigned equal weight.
We have assumed ${\rm log}~g = 0.8$ for all And II
program stars; the expected variations in surface gravity have a negligible
effect on the derived metallicity (see, $e.g.$, Figure 8-12).
For the five different indices, we find best-fit mean metallicities
of $-1.42$ (G4300), $-2.16$ (Mg$_1$), $-1.54$ (Mg$_2$), $-1.90$ (Mg$b$) 
and $-1.33$ (Fe5406) dex. The average of these five values is
${\overline{\rm [Fe/H]}} = -1.67\pm0.16$ dex where the quoted uncertainty
refers to the error in the mean. This is 
our best estimate of the mean metallicity of the And II stars
for we have LRIS spectra. By comparison, if the measurements are weighted according to 
the inverse square of the index uncertainties, the respective metallicities are 
$-1.67$, $-2.34$, $-1.62$, $-1.65$ and $-1.60$ dex, with a mean of 
${\overline{\rm [Fe/H]}} = -1.78\pm0.14$ dex.

The above estimate of ${\overline{\rm [Fe/H]}} = -1.67\pm0.16$ dex is biased since, 
as is evident in Figure 3, the most metal-poor giants are brighter than their 
metal-rich counterparts and will thus be
preferentially included in our spectroscopic sample. The magnitude of this bias 
is easily calculated, however, by comparing the distribution in the $I$, $V-I$ CMD of the red giants
with LRIS spectra to that of the entire sample of And II red giants. Figure 13 compares the
distributions over the range $20.2 \le I \le 21.2$: $i.e.$, the upper one magnitude
of the RGB. For the photometric sample, we consider
only those stars within one arcminute of the galaxy's center in order to minimize 
contamination foreground stars and unresolved galaxies. 
The lower panel Figure 13 shows the relative number of stars located in seven ``bins"
defined by the globular cluster fiducial sequences shown in the upper panel
(ie, bin 1 refers to the region blueward of the leftmost curve, bin 2 refers to 
the region between the two leftmost curves, etc). Based on the distributions in the
given in lower panel, we find that the mean metallicity of our spectroscopic sample
is $0.2\pm0.1$ dex more metal-poor than the unbiased sample. Thus, our corrected
estimate for the mean metallicity of And II is
$${\overline{\rm [Fe/H]}} = -1.47\pm0.19~{\rm dex}.$$
This value is in close
agreement with the value of ${\overline{\rm [Fe/H]}} = -1.59^{+0.44}_{-0.12}$ dex
found by KNMF and is consistent with the somewhat higher value of 
${\overline{\rm [Fe/H]}} = -1.35\pm0.3$ dex
found in \S 4.1 using the dereddened color of the galaxy's giant branch.

Can the spread in metallicity inferred from the wide giant
giant branch evident in Figures 3 and 4 be confirmed
spectroscopically? The uncertainties of the Mg$b$ indices are sufficiently small
that it is possible to test this claim. Fitting a Gaussian to the
residuals between the measured  Mg$b$ indices and the best-fit relation shown in Figure 11
gives a dispersion of $\sigma({\rm Mg}b)$ = $0.5\pm0.2$ \AA . From Gorgas et al. (1993),
we expect $d$Mg$b$/$d$[Fe/H] $\sim 1.7$ \AA\ dex$^{-1}$ over this range in color.
Thus, the observed dispersion in Mg$b$ corresponds to 
$\sigma({\rm{[Fe/H]}}) = 0.29\pm0.12$ dex, which when combined with
the findings of section \S 4.1, give
$$\sigma({\rm{[Fe/H]}}) = 0.35\pm0.10~{\rm dex}$$
which we adopt as our best estimate for the intrinsic dispersion of And II.

\section{Comparison to Other Local Groups Dwarfs}

In the upper panel of Figure 14 we show the dependence of ${\overline{\rm [Fe/H]}}$
on galaxy magnitude for Local Group dE, dSph and ``dIrr/dSph transition" objects. The 
data are taken from Mateo (1998), supplemented by new observations from 
Armandroff et al. (1998, 1999), Grebel \& Guhathakurta (1999) and Caldwell (1999). 
The measured metallicity of
And II is consistent with that expected on the basis of its absolute
magnitude and the well-known metallicity-luminosity relation for dwarf galaxies
(Aaronson 1986; Caldwell et al. 1992). As the
lower panel of Figure 14 demonstrates, And II exhibits a 
spread in metallicity which is comparable in size to that that found from 
spectroscopy of Galactic dSphs and photometry of several dSph companions of M31.
The sample mean, $\sigma({\rm{[Fe/H]}})$ = 0.37$\pm$0.03 dex,
is indicated by the dashed line.

Interestingly, available evidence seems to suggest that
although the {\sl mean} stellar metallicity in these galaxies --- whose
luminosities span a range of more than three orders of magnitude --- depends rather 
sensitively on absolute magnitude, the {\sl dispersion} in metallicity 
does not.\altaffilmark{3}\altaffiltext{3}{The range in {\sl mass} is a factor of $\sim$ 20
if these dwarfs are embedded in dark halos of mass $M \sim 2\times10^7M_{\odot}$, as proposed
by Mateo (1993; 1998).}
High-resolution spectroscopy of additional Galactic dSphs, such as that
presented by Shetrone et al. (1998) for Draco, and intermediate-resolution 
spectroscopy for an expanded sample of M31 dwarf galaxies
will help refine our understanding of the chemical 
enrichment histories of these galaxies.

\acknowledgments
 
The authors thank Christian K\"onig for sending us his Thuan-Gunn photometry and Ed Olszewski
for providing coordinates of the four stars observed by Aaronson et al. (1985).
P.C. acknowledges support provided by the Sherman M. Fairchild Foundation. The research of J.G.C. is
supported, in part, by NSF grant 96-16729.
The W. M. Keck Observatory is operated as a scientific partnership between the California Institute of
Technology, the University of California, and the National Aeronautics and Space Administration. It was
made possible by the generous financial support of the W. M. Keck Foundation.

\begin{deluxetable}{llllllcll}                        
\scriptsize
\tablecolumns{9}                        
\tablewidth{0pc}                        
\tablecaption{Candidate Red Giants in Andromeda II Observed With LRIS\label{tbl-2}}                    
\tablehead{                        
\colhead{ID} &                       
\colhead{$\alpha$(2000)} &                       
\colhead{$\delta$(2000)} &
\colhead{$V$} &                      
\colhead{$(V-I)$} &                       
\colhead{$g$} &
\colhead{$(g-r)$} &
\colhead{KNMF}  &                     
\colhead{Comments} \nl
\colhead{} &
\colhead{(h:m:s)} &
\colhead{($\circ$:$\prime$:$\prime\prime$)} &
\colhead{(mag)} &
\colhead{(mag)} &
\colhead{(mag)} &
\colhead{(mag)} &
\colhead{}  &
\colhead{} 
}                        
\startdata                        
1  & 01:16:22.11 & 33:25:59.2 &22.26 & 1.36 & 21.84 & 0.72 & NE-382 & member \nl
2  & 01:16:24.19 & 33:25:51.0 &22.42 & 1.75 & 21.80 & 0.94 & NE-228 & member \nl
3  & 01:16:26.52 & 33:25:41.8 &22.33 & 1.92 & 22.00 & 0.92 & NE-062 & member? carbon star \nl
4  & 01:16:25.19 & 33:25:31.9 &21.91 & 1.62 & 21.48 & 0.80 & NE-161 & member \nl
5  & 01:16:21.77 & 33:25:06.2 &21.89 & 1.72 & 21.64 & 0.82 & SE-227 & member \nl
6  & 01:16:21.16 & 33:24:57.6 &22.37 & 1.44 & 23.00 & 0.37 & SE-267 & member \nl
7  & 01:16:22.80 & 33:24:48.7 &22.29 & 1.67 & 22.02 & 0.76 & SE-185 & member \nl
8  & 01:16:26.96 & 33:24:39.4 &22.17 & 1.93 & 21.69 & 1.08 & SE-041 & member \nl
9  & 01:16:27.34 & 33:24:30.6 &22.43 & 2.50 & 21.93 & 1.12 & SE-034 & non-member \nl
10 & 01:16:25.63 & 33:24:21.3 &22.30 & 1.51 & 22.27 & 0.35 & SE-084 & member \nl
11$^{\rm a}$ & 01:16:25.24 & 33:24:11.5 &21.94 & 1.63 & 21.74 & 0.64 & SE-098 & member \nl
12 & 01:16:27.34 & 33:24:00.1 &22.25 & 2.01 & 22.00 & 0.73 & SE-035 & member \nl
13 & 01:16:24.76 & 33:23:40.7 &22.19 & 1.75 & 22.02 & 0.97 & SE-053 & member \nl
14 & 01:16:23.50 & 33:23:26.7 &22.03 & 2.53 & 21.06 & 1.40 & SE-153 & member \nl
15 & 01:16:20.68 & 33:23:18.1 &22.02 & 1.52 & 21.59 & 0.91 & SE-289 & member \nl
16 & 01:16:19.33 & 33:23:07.2 &22.19 & 1.72 & 21.76 & 1.17 & SE-319 & member \nl
17 & 01:16:22.73 & 33:22:56.2 &21.96 & 1.50 & 21.63 & 0.74 & SE-188 & member \nl
18 & 01:16:24.93 & 33:22:45.8 &21.69 & 1.65 & 21.39 & 0.68 & SE-104 & member \nl
19 & 01:16:27.35 & 33:22:23.0 &21.67 & 1.38 &       &      &        & member \nl
20$^{\rm a}$ & 01:16:23.56 & 33:21:49.1 &22.20 & 1.36 & 22.05 & 0.43 & SE-151 & member \nl
21 & 01:16:21.87 & 33:21:37.2 &21.89 & 1.98 &       &      &        & non-member \nl
22 & 01:16:21.96 & 33:21:14.4 &21.65 & 1.69 &       &      &        & member \nl
23 & 01:16:20.79 & 33:21:03.0 &22.29 & 1.36 & 22.70 &-0.07 & SE-278 & member \nl
24 & 01:16:24.66 & 33:20:28.3 &21.52 & 1.31 &       &      &        & non-member \nl
25 & 01:16:19.17 & 33:20:06.6 &22.47 & 1.53 &       &      &        & member \nl
26 & 01:16:22.27 & 33:19:50.9 &21.82 & 1.66 &       &      &        & member \nl
27 & 01:16:24.78 & 33:19:26.4 &22.26 & 1.38 &       &      &        & low S/N; member \nl
28 & 01:16:22.10 & 33:18:59.3 &21.77 & 1.65 &       &      &        & member \nl
29 & 01:16:22.02 & 33:26:02.2 &22.15 & 1.49 & 22.56 & 0.95 & NE-385 & member \nl
30 & 01:16:26.55 & 33:25:48.6 &21.37 & 1.65 & 21.89 & 0.85 & NE-061 & member \nl
31 & 01:16:19.54 & 33:25:32.4 &22.49 & 1.11 & 22.48 & 0.30 & NE-552 & low S/N; membership uncertain  \nl
32 & 01:16:25.31 & 33:25:10.5 &21.82 & 1.81 &       &      &        & member \nl
33 & 01:16:23.67 & 33:24:59.3 &22.16 & 1.72 & 22.51 & 0.70 & SE-146 & member \nl
34 & 01:16:19.86 & 33:24:50.1 &22.19 & 1.48 &       &      &        & member \nl
35 & 01:16:20.77 & 33:24:37.1 &22.14 & 1.46 & 22.57 & 0.67 & SE-276 & member \nl
36 & 01:16:27.49 & 33:24:25.8 &22.05 & 1.78 & 22.45 & 0.63 & SE-032 & member \nl
37 & 01:16:21.17 & 33:24:02.5 &22.03 & 1.56 & 22.45 & 0.82 & SE-259 & member \nl
38 & 01:16:23.07 & 33:23:42.0 &22.61 & 1.56 & 22.68 & 0.70 & SE-172 & emission-line gal., $z$ = 0.55 \nl
39 & 01:16:23.16 & 33:23:23.3 &22.20 & 1.88 & 22.94 & 1.30 & SE-166 & member \nl
40 & 01:16:26.55 & 33:23:08.6 &22.52 & 1.76 & 22.99 & 0.97 & SE-053 & member \nl 
41 & 01:16:19.43 & 33:22:56.1 &22.84 & 2.70 & 23.52 & 1.42 & SE-317 & non-member \nl
42 & 01:16:24.25 & 33:22:40.2 &22.40 & 1.41 & 22.72 & 0.54 & SE-129 & member \nl
43 & 01:16:19.70 & 33:21:38.8 &22.51 & 1.47 &       &      &        & member \nl
44 & 01:16:22.47 & 33:21:20.3 &22.79 & 1.66 &       &      &        & member \nl
45 & 01:16:20.15 & 33:21:09.0 &21.82 & 1.49 &       &      &        & member \nl
46 & 01:16:26.92 & 33:20:31.9 &22.69 & 2.18 &       &      &        & low S/N; membership uncertain \nl
47 & 01:16:23.97 & 33:20:14.8 &22.45 & 1.36 &       &      &        & member \nl
48 & 01:16:20.26 & 33:19:56.6 &21.78 & 1.53 &       &      &        & member \nl
49 & 01:16:21.92 & 33:19:31.8 &22.04 & 1.93 &       &      &        & member \nl
50 & 01:16:22.26 & 33:19:01.3 &22.12 & 1.50 &       &      &        & member \nl
\enddata                        
\clearpage

\tt{a}{Object included on both LRIS masks.}

\end{deluxetable}                        
\clearpage

\begin{deluxetable}{rrrrrr}
\tablecolumns{6}
\tablewidth{0pc}
\tablecaption{Lick/IDS Standard Stars Observed With LRIS\label{tbl-2}}
\tablehead{
\colhead{Cluster} &
\colhead{Star} &
\colhead{$M_V$} &
\colhead{$(V-K)_0$} &
\colhead{$(V-I)_0$} &
\colhead{[Fe/H]} \nl
\colhead{} &
\colhead{} &
\colhead{(mag)} &
\colhead{(mag)} &
\colhead{(mag)} &
\colhead{(dex)}
}
\startdata
M13 &B786       &$-$2.23 &3.57 &1.50 &$-$1.54 \nl
    &B818       &$-$0.29 &1.98 &0.86 &$-$1.54 \nl
    &I-2$^a$    &$-$0.06 &2.27 &0.97 &$-$1.54 \nl
    &I-24$^a$   &$-$1.46 &2.72 &1.14 &$-$1.54 \nl
    &I-48$^a$   &$-$2.27 &3.47 &1.46 &$-$1.54 \nl
    &II-67$^a$  &$-$2.20 &3.54 &1.49 &$-$1.54 \nl
    &II-76$^a$  &$-$1.80 &2.95 &1.23 &$-$1.54 \nl
    &II-90$^a$  &$-$2.09 &3.49 &1.47 &$-$1.54 \nl
    &III-72$^a$ &   0.81 &2.00 &0.87 &$-$1.54 \nl
    &IV-25$^a$  &$-$2.23 &3.43 &1.44 &$-$1.54 \nl
M92 &III13&$-$2.57 &3.03&1.28&$-$2.29 \nl
    &IV114&$-$0.77 &2.35&1.03&$-$2.29 \nl
M71 &I-64 &$-$0.30 &2.88&1.21&$-$0.73 \nl
    &A2   &   1.13 &2.19&0.93&$-$0.73 \nl
M67 &F094 &   3.33 &1.21&0.58&$-$0.05 \nl
    &F105 &   0.80 &2.66&1.14&$-$0.05 \nl
    &F115 &   3.15 &1.30&0.61&$-$0.05 \nl
    &F117 &   3.11 &1.86&0.82&$-$0.05 \nl
    &F125 &   4.36 &1.23&0.58&$-$0.05 \nl
    &IV20 &   1.71 &2.44&1.05&$-$0.05 \nl
\enddata
\clearpage

\tt{a}{Star not included in Gorgas et al. (1993) or Worthey et al. (1994). $(V-K)_0$ photometry
taken from Cohen, Persson \& Frogel (1978).}

\end{deluxetable}                        
\clearpage
 
\begin{deluxetable}{rrrrrrr}
\tablecolumns{7}
\scriptsize
\tablewidth{0pc}
\tablecaption{Index Measurements for Andromeda II Members \label{tbl-2}}
\tablehead{
\colhead{ID} &
\colhead{G4300} &
\colhead{Mg$_1$} &
\colhead{Mg$_2$} &
\colhead{Mg$b$} &
\colhead{Fe5406} &
\colhead{Comments} \nl
\colhead{ } &
\colhead{(\AA )} &
\colhead{(mag)} &
\colhead{(mag)} &
\colhead{(\AA )} &
\colhead{(\AA )} &
\colhead{ } 
}
\startdata
 1 &  6.16$\pm$0.50 &  0.033$\pm$0.046 &  0.083$\pm$0.047 & 1.23$\pm$0.52 & 0.25$\pm$0.05 & \nl
 2 &                &  0.080$\pm$0.057 &  0.276$\pm$0.055 & 1.52$\pm$0.33 & 2.46$\pm$0.33 & \nl
 4 &  5.37$\pm$0.71 &  0.071$\pm$0.039 &  0.153$\pm$0.041 & 0.75$\pm$0.36 & 2.49$\pm$0.21 & \nl
 5 &  5.79$\pm$0.10 &  0.088$\pm$0.043 &  0.164$\pm$0.045 & 0.19$\pm$0.10 & 2.59$\pm$0.37 & \nl
 6 &  6.26$\pm$0.68 &  0.023$\pm$0.046 &  0.066$\pm$0.045 & 0.93$\pm$0.38 & 1.11$\pm$0.24 & \nl
 7 &                &  0.112$\pm$0.050 &  0.202$\pm$0.051 & 1.43$\pm$0.13 & 2.66$\pm$0.07 & \nl
 8 &                &  0.163$\pm$0.049 &  0.249$\pm$0.045 & 1.62$\pm$0.18 & 3.24$\pm$0.15 & \nl
10 &  8.99$\pm$2.65 &  0.062$\pm$0.042 &  0.097$\pm$0.046 & 0.21$\pm$0.39 & 1.61$\pm$0.04 & \nl
11 &  7.97$\pm$1.11 &  0.066$\pm$0.039 &  0.100$\pm$0.041 & 0.38$\pm$0.67 & 1.41$\pm$0.11 &Mask 1\nl
   &  7.15$\pm$1.17 &  0.076$\pm$0.043 &  0.118$\pm$0.042 & 1.35$\pm$0.05 & 2.35$\pm$0.36 &Mask 2\nl
   &  7.65$\pm$0.81 &  0.071$\pm$0.029 &  0.109$\pm$0.029 & 1.34$\pm$0.05 & 1.49$\pm$0.11 &Average\nl
12 & 11.42$\pm$0.61 &  0.120$\pm$0.047 &  0.226$\pm$0.051 & 0.19$\pm$0.76 & 3.05$\pm$0.05 & \nl
13 &  6.38$\pm$0.09 &  0.090$\pm$0.046 &  0.216$\pm$0.046 & 2.67$\pm$0.67 & 3.49$\pm$0.26 & \nl
14 & 10.98$\pm$2.51 &  0.100$\pm$0.042 &  0.431$\pm$0.049 & 5.74$\pm$1.52 & 3.01$\pm$0.07 & \nl
15 &  7.24$\pm$0.16 &  0.006$\pm$0.041 &  0.063$\pm$0.040 & 1.08$\pm$0.12 & 0.40$\pm$0.15 & \nl
16 &                &  0.098$\pm$0.049 &  0.176$\pm$0.048 & 1.29$\pm$0.85 & 2.41$\pm$0.44 & \nl
17 &  7.60$\pm$0.85 &  0.038$\pm$0.041 &  0.071$\pm$0.039 & 0.16$\pm$0.07 & 1.47$\pm$0.20 & \nl
18 &  5.10$\pm$0.47 &  0.039$\pm$0.034 &  0.082$\pm$0.035 & 0.85$\pm$0.73 & 0.91$\pm$0.05 & \nl
19 &  7.47$\pm$1.01 &  0.050$\pm$0.034 &  0.090$\pm$0.033 & 0.58$\pm$0.42 & 1.51$\pm$0.09 & \nl
20 &  4.31$\pm$0.09 &  0.019$\pm$0.044 &  0.050$\pm$0.041 & 0.40$\pm$0.58 & 0.43$\pm$0.13 &Mask 1\nl
   &  4.46$\pm$0.58 &  0.012$\pm$0.046 &  0.087$\pm$0.049 & 0.95$\pm$0.68 &               &Mask 2\nl
   &  4.31$\pm$0.09 &  0.016$\pm$0.032 &  0.065$\pm$0.031 & 0.63$\pm$0.44 & 0.43$\pm$0.13 &Average\nl
22 &  2.64$\pm$1.06 &  0.082$\pm$0.037 &  0.145$\pm$0.038 & 1.06$\pm$1.10 & 2.03$\pm$0.34 & \nl
23 &  5.03$\pm$0.32 &  0.016$\pm$0.047 &  0.078$\pm$0.047 & 1.11$\pm$0.28 & 0.53$\pm$0.05 & \nl
25 &                &  0.076$\pm$0.061 &  0.136$\pm$0.063 & 2.19$\pm$0.31 & 1.05$\pm$0.08 & \nl
26 &                &  0.125$\pm$0.055 &  0.160$\pm$0.052 & 0.62$\pm$0.32 & 3.03$\pm$0.11 & \nl
28 &                &  0.125$\pm$0.069 &  0.168$\pm$0.067 & 2.31$\pm$0.08 & 0.97$\pm$0.40 & \nl
29 &                &  0.065$\pm$0.055 &  0.138$\pm$0.057 & 0.89$\pm$0.34 & 1.00$\pm$0.06 & \nl
30 &                &  0.088$\pm$0.048 &  0.101$\pm$0.047 &               & 1.98$\pm$0.29 & \nl
32 &                &  0.064$\pm$0.044 &  0.206$\pm$0.045 & 3.42$\pm$0.05 & 2.20$\pm$0.56 & \nl
33 &                &  0.091$\pm$0.053 &  0.176$\pm$0.055 & 1.26$\pm$0.67 & 3.16$\pm$0.06 & \nl
34 &  5.66$\pm$2.38 &  0.039$\pm$0.047 &  0.069$\pm$0.049 & 0.52$\pm$0.30 & 0.54$\pm$0.18 & \nl
35 &                &  0.134$\pm$0.053 &  0.202$\pm$0.052 & 1.32$\pm$0.13 & 2.38$\pm$0.63 & \nl
36 &                &  0.066$\pm$0.049 &  0.115$\pm$0.049 &               & 2.39$\pm$0.10 & \nl
37 &  4.98$\pm$0.28 &  0.029$\pm$0.043 &  0.078$\pm$0.043 & 0.22$\pm$0.56 & 0.36$\pm$0.83 & \nl
39 &                &  0.172$\pm$0.053 &  0.265$\pm$0.057 & 2.65$\pm$0.65 & 3.25$\pm$0.07 & \nl
40 &                &  0.036$\pm$0.060 &  0.268$\pm$0.060 & 1.26$\pm$0.46 & 2.62$\pm$0.63 & \nl
42 &  6.81$\pm$0.37 &  0.070$\pm$0.057 &  0.107$\pm$0.056 & 1.84$\pm$0.55 & 1.11$\pm$0.74 & \nl
43 &  1.93$\pm$0.09 &  0.008$\pm$0.055 &  0.042$\pm$0.059 &               & 2.13$\pm$0.36 & \nl
44 &  6.10$\pm$0.09 &  0.091$\pm$0.041 &  0.151$\pm$0.041 & 1.19$\pm$0.40 & 0.96$\pm$0.50 & \nl
45 &  5.78$\pm$0.39 &  0.018$\pm$0.042 &  0.061$\pm$0.041 & 1.27$\pm$0.63 & 1.42$\pm$0.08 & \nl
47 &  2.40$\pm$1.21 &  0.029$\pm$0.054 &  0.090$\pm$0.053 &               & 0.32$\pm$3.48 & \nl
48 &  5.40$\pm$0.08 &  0.017$\pm$0.043 &  0.096$\pm$0.044 & 1.38$\pm$0.67 & 0.86$\pm$0.04 & \nl
49 &  2.90$\pm$1.28 &  0.136$\pm$0.049 &  0.232$\pm$0.051 & 2.77$\pm$0.35 & 0.63$\pm$0.08 & \nl
50 &                &                  &  0.055$\pm$0.063 & 2.43$\pm$0.82 & 0.86$\pm$0.18 & \nl
\enddata
\clearpage
\end{deluxetable}

\clearpage

\clearpage
 
\plotone{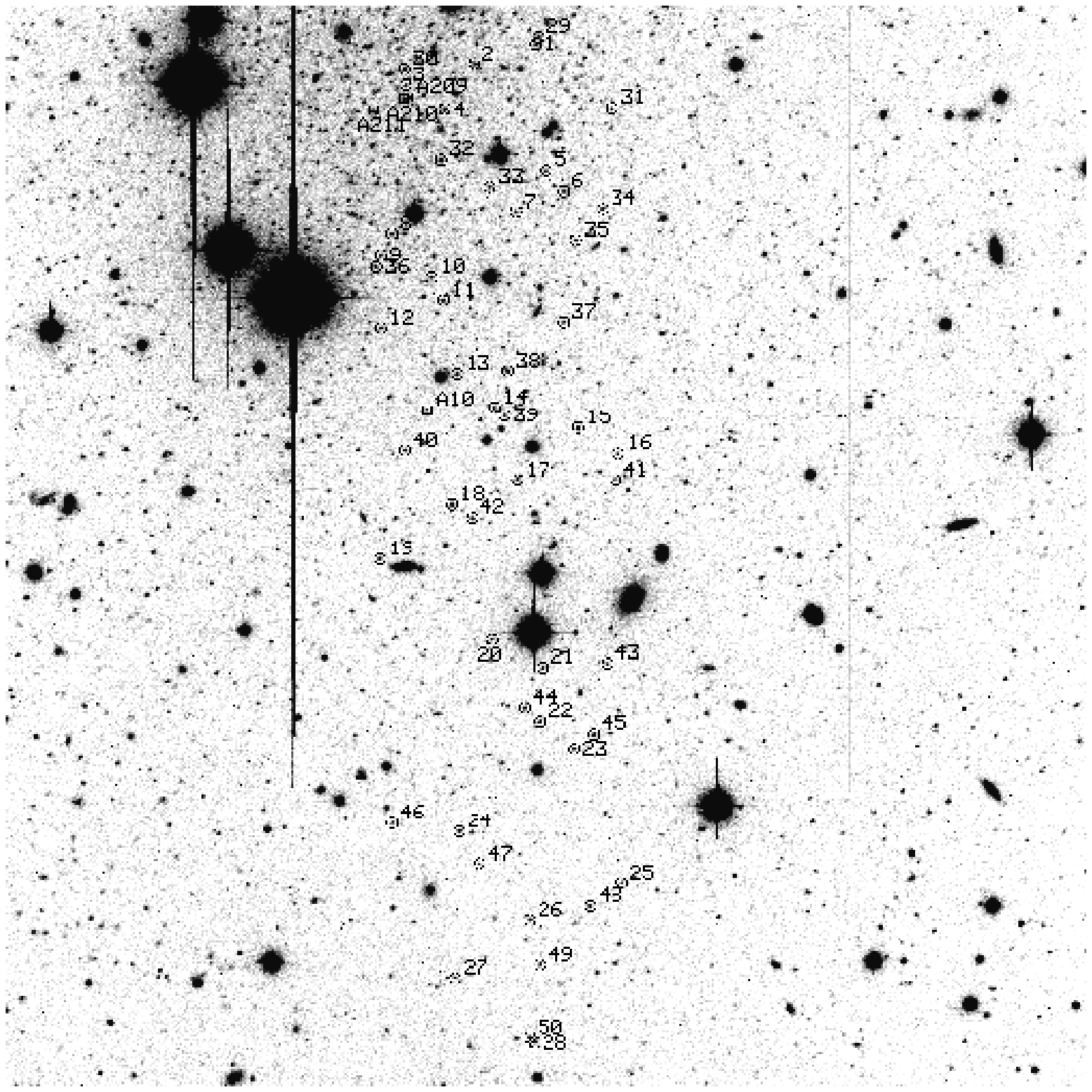}

\figcaption[and2b.01.ps]{Finding chart for Andromeda II red giant candidates observed with 
LRIS in multi-slit mode (circles). The numbering scheme corresponds to that given in Table 1. 
North is to the top and east is to the left on this $V$-band COSMIC image, which measures 
7\farcm6$\times$7\farcm6. The four stars observed by Aaronson et al. (1985) are indicated
by the squares.
\label{fig1}}

\clearpage

\plotone{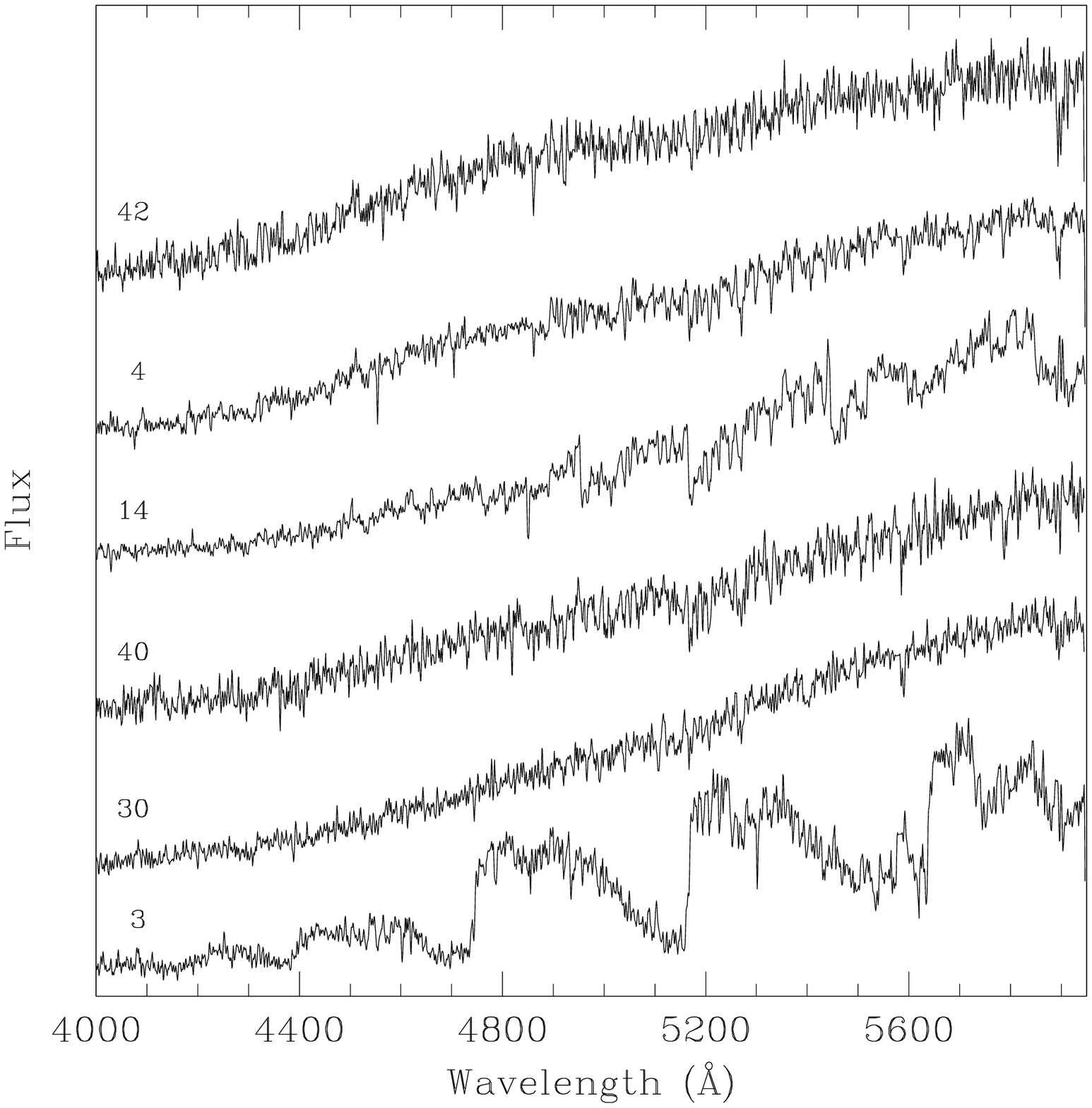}

\figcaption[and2b.02.ps]{A representative sample of LRIS spectra for Andromeda II giants.
The spectra shown here correspond to program stars which span a wide range in magnitude
and color (see Table 1). At 5000 \AA , the spectra have 30 $\lae$ S/N $\lae$ 50 per
resolution element. Note the obvious Swan C$_2$ bands in the spectrum of star \#3.
\label{fig2}}

\clearpage

\plotone{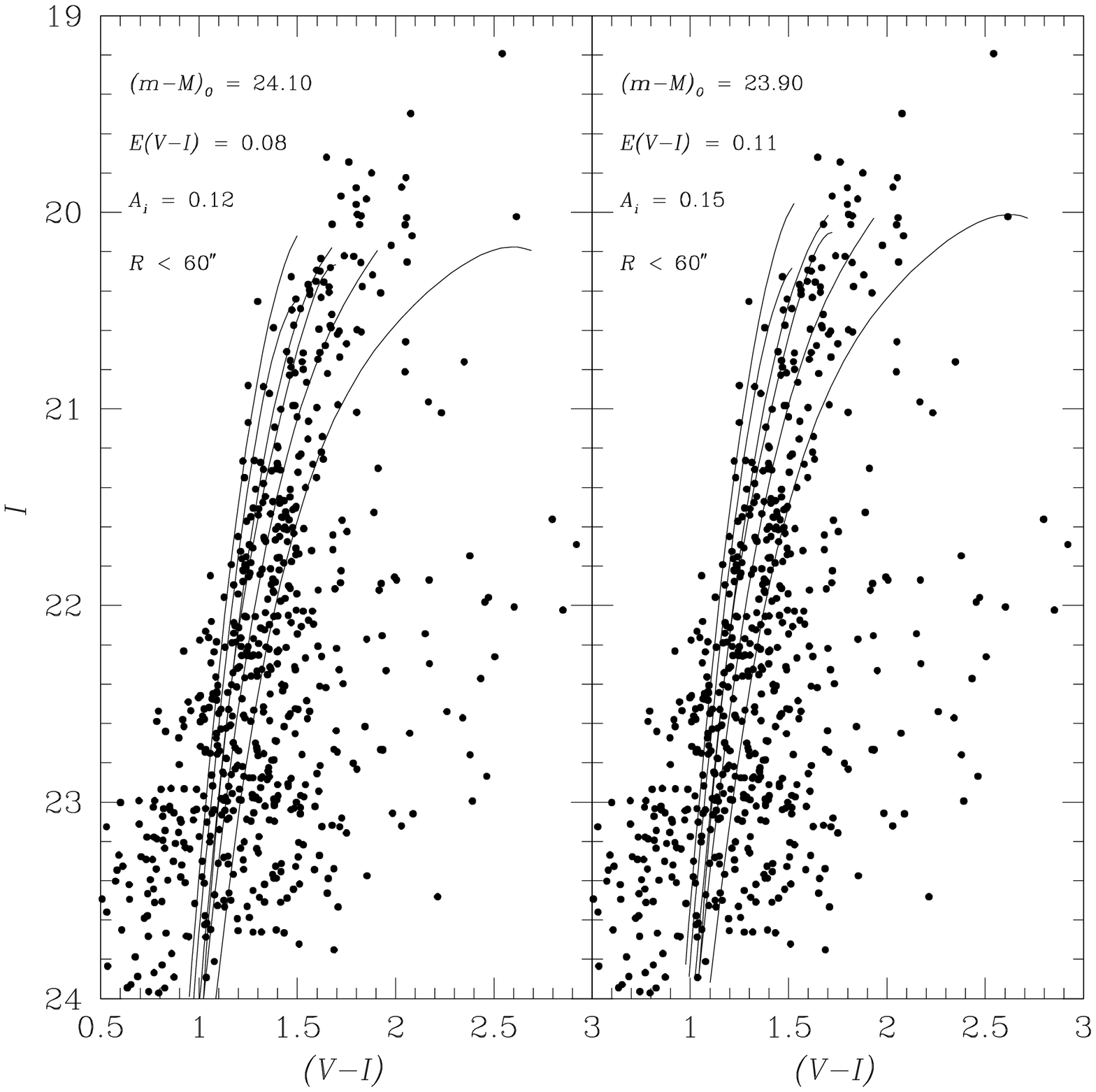}

\figcaption[and2b.03.ps]{$I,(V-I)$ color-magnitude diagram for Andromeda II. In both panels,
only unresolved objects within 60$^{\prime\prime}$ of the center of the galaxy are shown,
along with fiducial sequences for the red giant branches of six Galactic globular clusters:
from left to right, the solid lines are M15 ([Fe/H] = $-$2.22), NGC 6397 ($-$1.95),
M2 ($-$1.62), NGC 6752 ($-$1.55), NGC 1851 ($-$1.26) and 47 Tuc ($-$0.76). In the left
panel, we show the fiducial sequences shifted by {\it E(V-I)} = 0.08, $A_V$ = 0.12 and 
$(m-M)_0$ = 24.10. This reddening corresponds to the value of {\it E(B-V)} = 0.06$\pm$0.01
indicated by the reddening maps of Schlegel, Finkbeiner \& Davis (1998). The right
panel show the same sequences shifted by {\it E(V-I)} = 0.11, $A_V$ = 0.15 and
$(m-M)_0$ = 23.90, equivalent to adopting {\it E(B-V)} = 0.08.
\label{fig3}}

\clearpage

\plotone{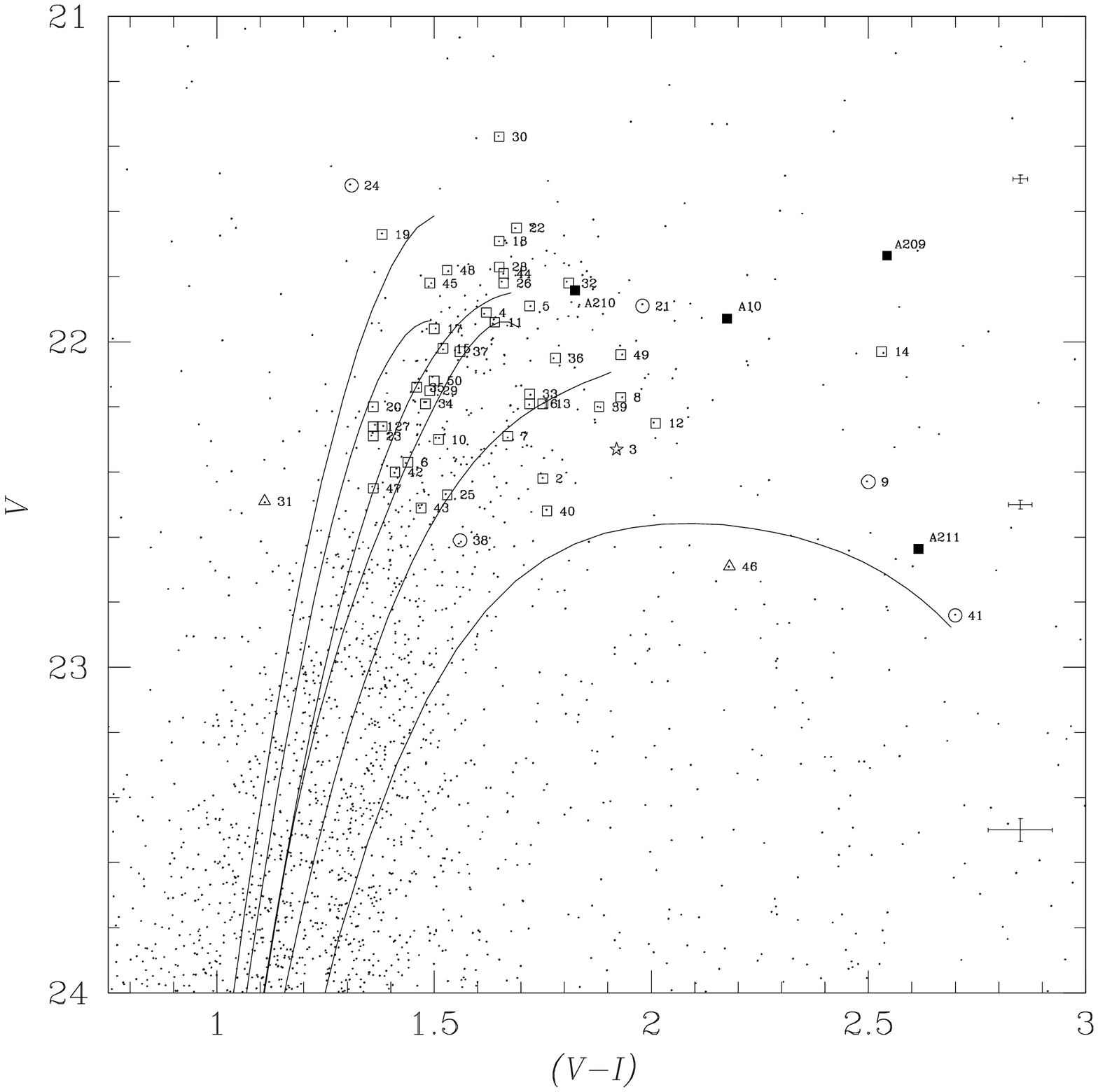}

\figcaption[and2b.04.ps]{$V,(V-I)$ color-magnitude diagram for Andromeda II.
The globular cluster fiducial sequences are the same as in Figure 2. We have adopted
{\it E(V-I)} = 0.08, $A_V$ = 0.19 and $(m-M)_0$ = 24.10. Open squares indicate member 
giants for which we have LRIS spectra, non-members are shown as circles and triangles 
indicate objects whose membership in Andromeda II is uncertain. The carbon star 
shown is indicated by the open star. 
The four stars studied by Aaronson et al. (1985) are indicated by the filled squares.
Typical internal photometric errors are shown at the right.
\label{fig4}}

\clearpage
\plotone{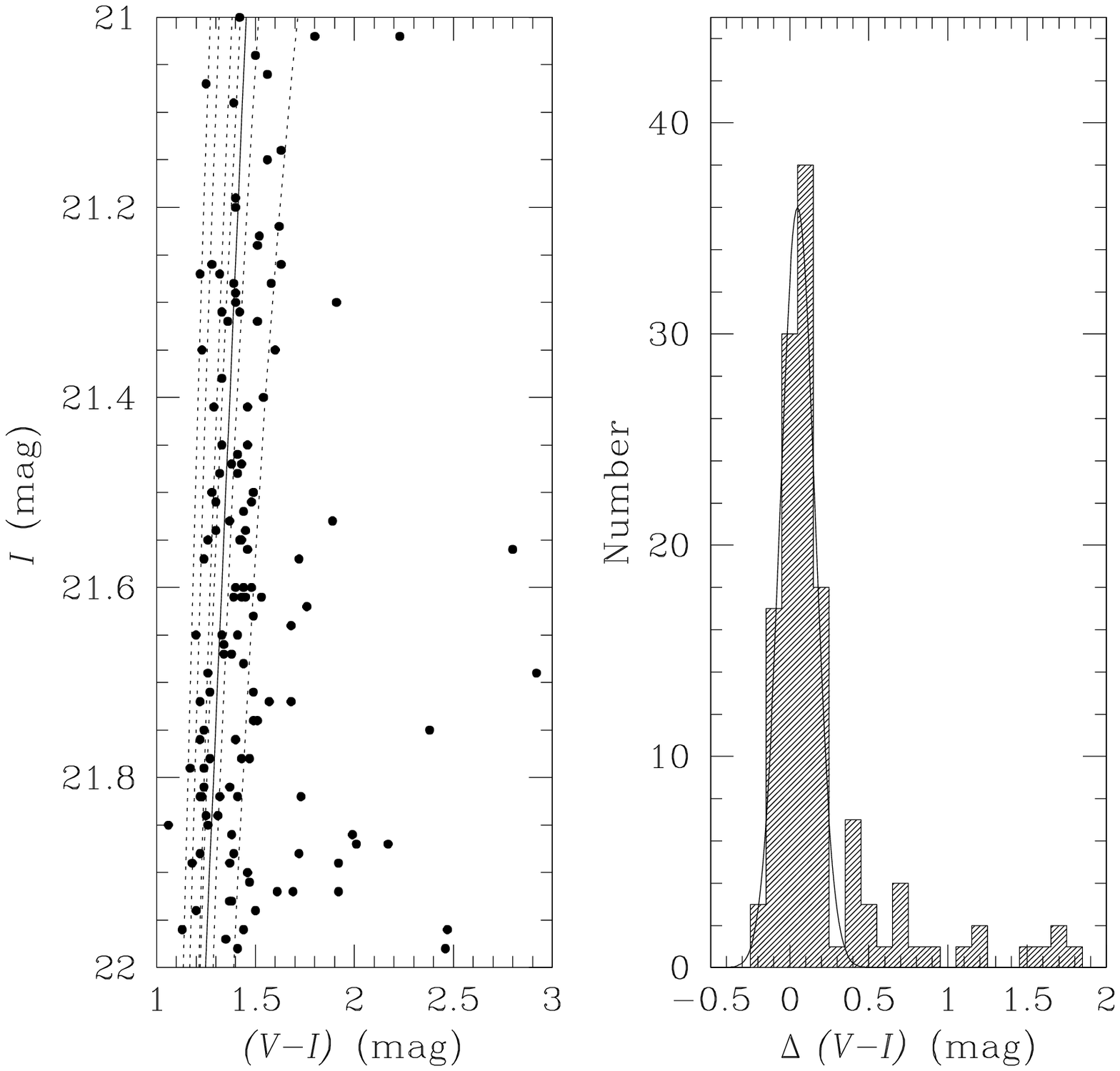}
\figcaption[and2b.05.ps]{(Left Panel) $I,(V-I)$ color-magnitude for all unresolved objects 
located within one arcminute of the center of Andromeda II and having $21 \le I \le 22$. 
The dotted curves show the globular cluster fiducial sequences shown in Figure 3. 
The solid line shows the adopted ridge line for Andromeda II; residuals about this
line are presented in the adjoining figure.
(Right Panel) Histogram of $(V-I)$ color residuals for the sample of unresolved 
objects shown in the previous panel. The solid curve is a Gaussian 
having dispersion ${\sigma}(V-I)$ = 0.11 mag. The mean internal photometric uncertainty 
for the stars in this interval is $\sigma(V-I$ = 0.05 mag. If the intrinsic
dispersion in color is ascribed entirely to variations in metallicity,
then $\sigma$([Fe/H]) $\simeq$ 0.46$\pm$0.17 dex (see text for details).
\label{fig5}}

\plotone{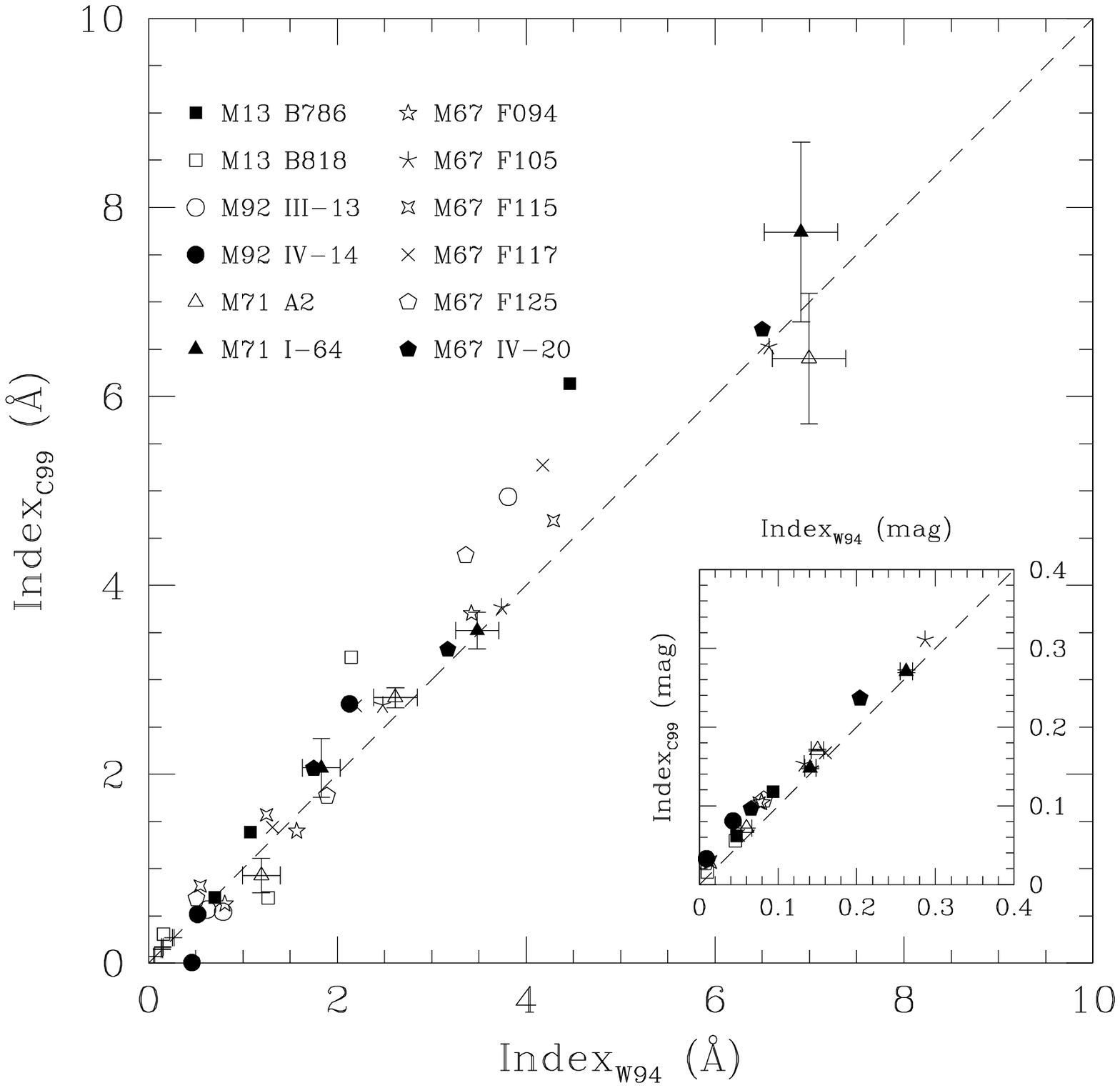}

\figcaption[and2b.06.ps]{(Main Panel) Comparison of the G4300, Mg$b$ and Fe5406 line 
indices measured from our LRIS spectra (C99) with those tabulated by Worthey et al. 
(1994; W94) for 12 Lick/IDS standards in the clusters M13, M71, M92 and M67. The dashed 
line shows the one-to-one relation. For clarity, errorbars are shown 
for the two giants in M71 only.
(Inset) Mg$_1$ and Mg$_2$ indices measured from our LRIS spectra compared to the published 
values from Worthey et al. (1994). The dashed line shows the one-to-one relation.
\label{fig6}}

\clearpage

\plotone{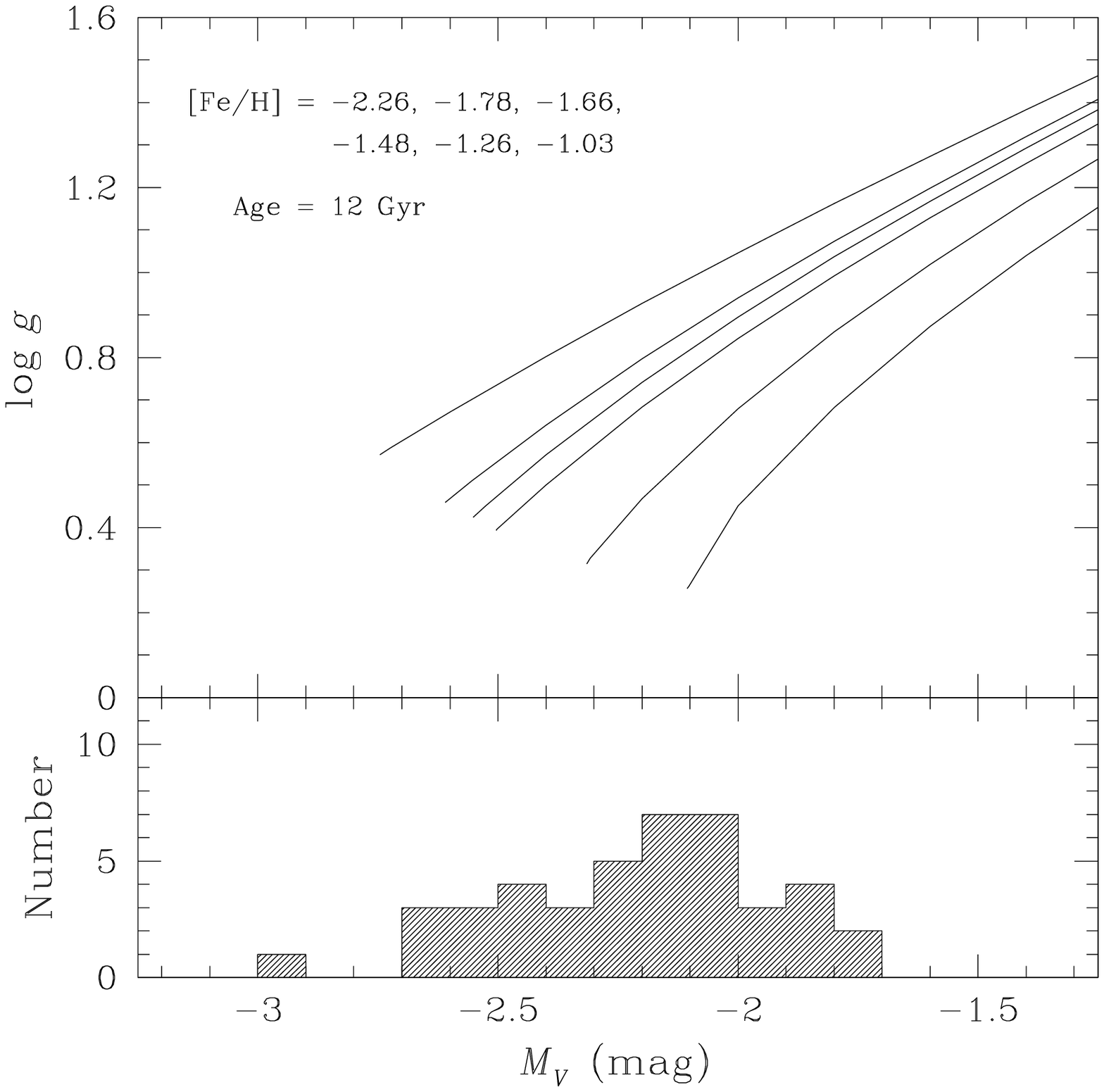}

\figcaption[and2b.07.ps]{(Upper Panel) Surface gravity plotted as a function of absolute
visual magnitude for metal-poor red giants of age 12 Gyr (Bergbusch \& VandenBerg 1992). 
From top to bottom, the curves indicate models having [Fe/H] = $-$2.26, $-$1.78, $-$1.66, 
$-$1.48, $-$1.26 and $-$1.03.
(Lower Panel) Histogram of absolute visual magnitudes for Andromeda II members having
measured line indices. Based on these models, the program stars are expected to have 
surface gravities in the range $0.4 \lae {\rm log}~g \lae 1.2$, with a mean value of
${\rm log}~g$ $\sim$ 0.8.
\label{fig7}}

\clearpage

\plotone{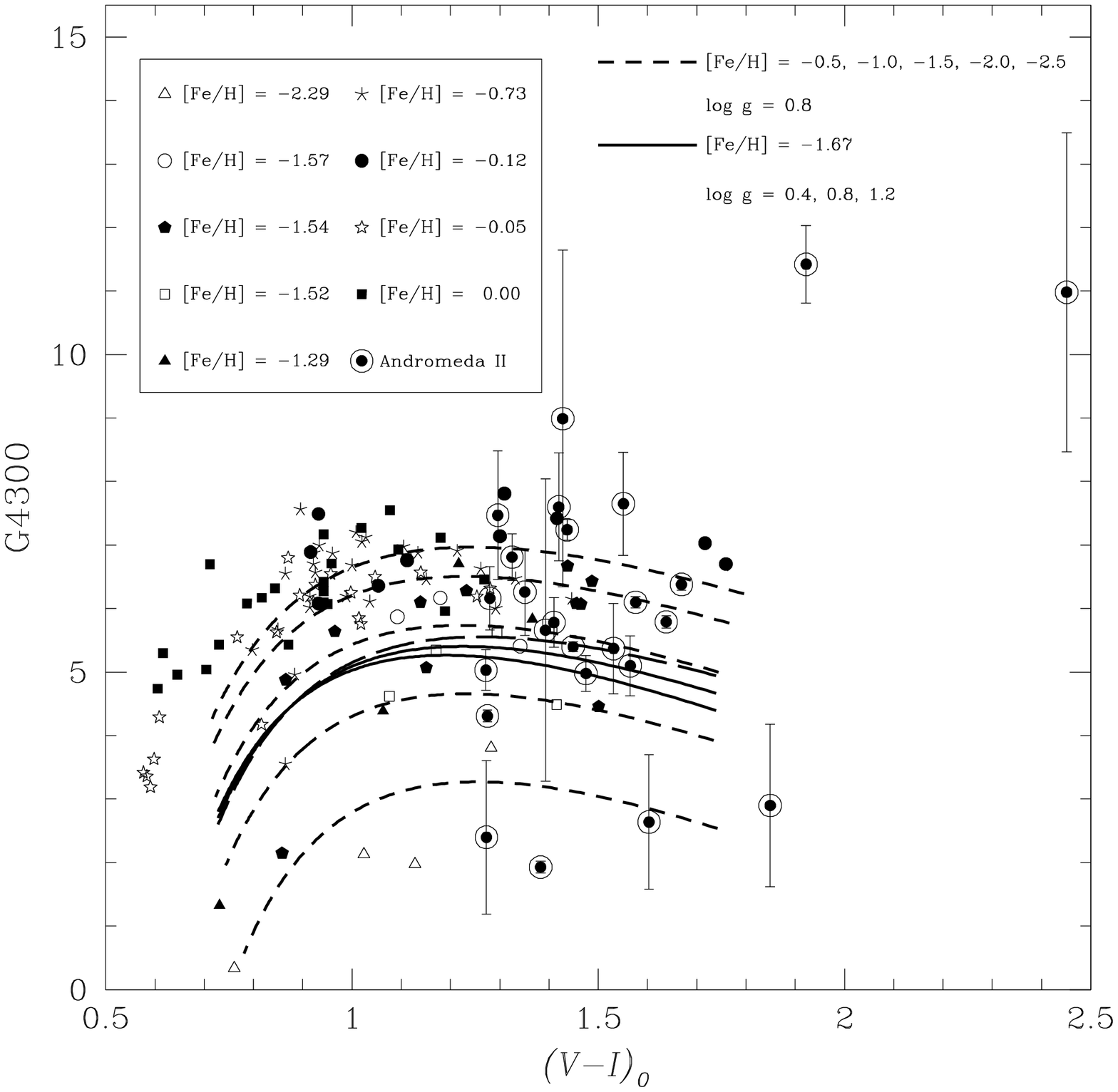}

\figcaption[and2b.08.ps]{G4300 line indices plotted against {\it (V-I)}$_0$ (see text) for 
Lick/IDS standard stars and Andromeda II red giants. The solid and dashed lines
show the calibration of Gorgas et al. (1993) for various metallicities and
surface gravities. Here [Fe/H] decreases from top to bottom for the dashed curves,
and ${\rm log}~g$ decreases from top to bottom at the red extrema for the solid 
curves. The indices for the standard stars are from Gorgas et al. (1993) and 
Worthey et al. (1994). The symbols correspond to individual stars in the following 
clusters: M92 ([Fe/H] = $-$2.29), M3 ($-$1.57), M13, ($-$1.54), M10 ($-$1.52), 
M5 ($-$1.29), M71 ($-$0.73), NGC 7789 ($-$0.12), M67 ($-$0.05) and NGC 188 (0.00).
\label{fig8}}

\clearpage

\plotone{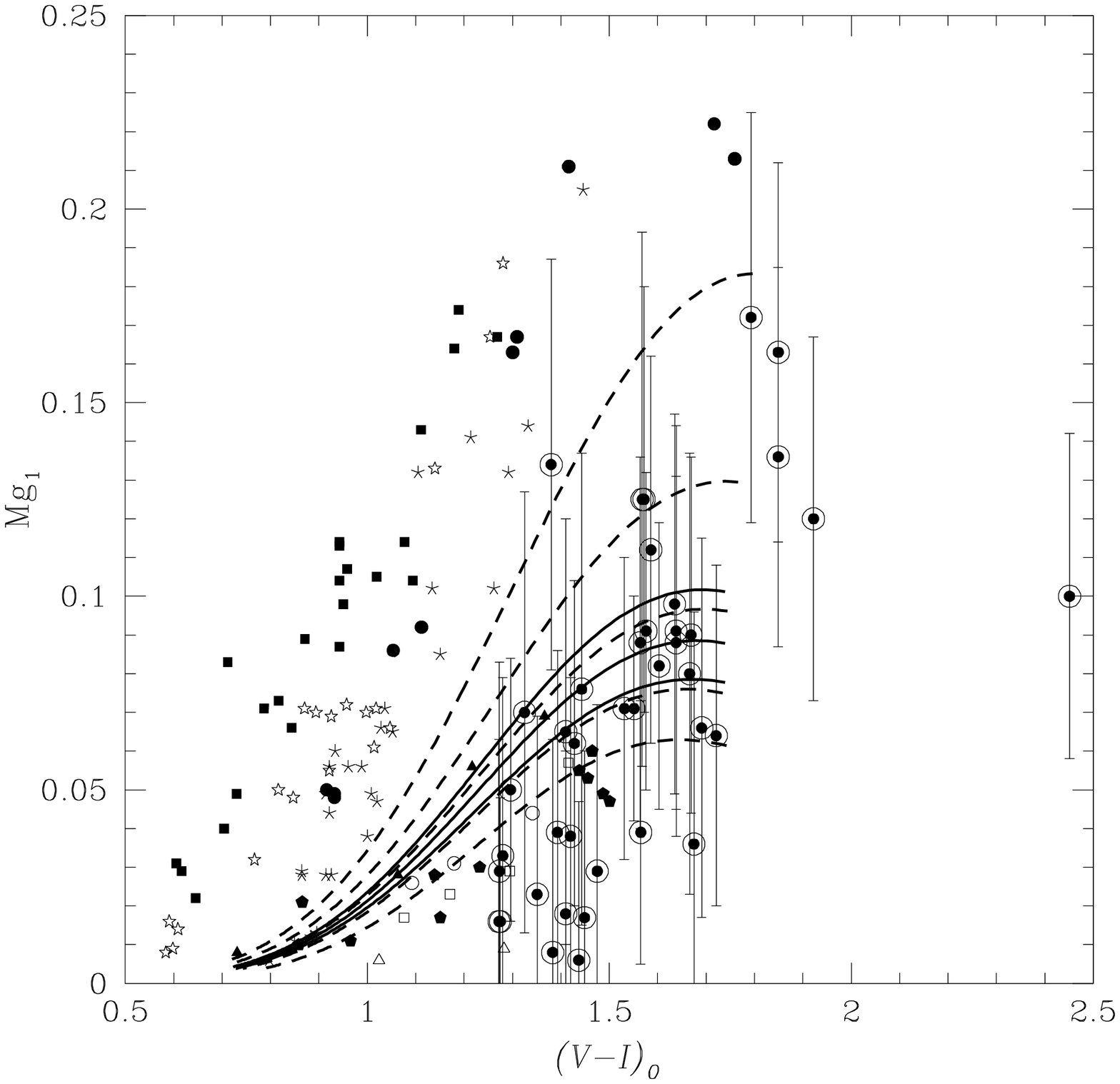}

\figcaption[and2b.09.ps]{Same as Figure 8, except for the Mg$_1$ index.
\label{fig9}}

\clearpage

\plotone{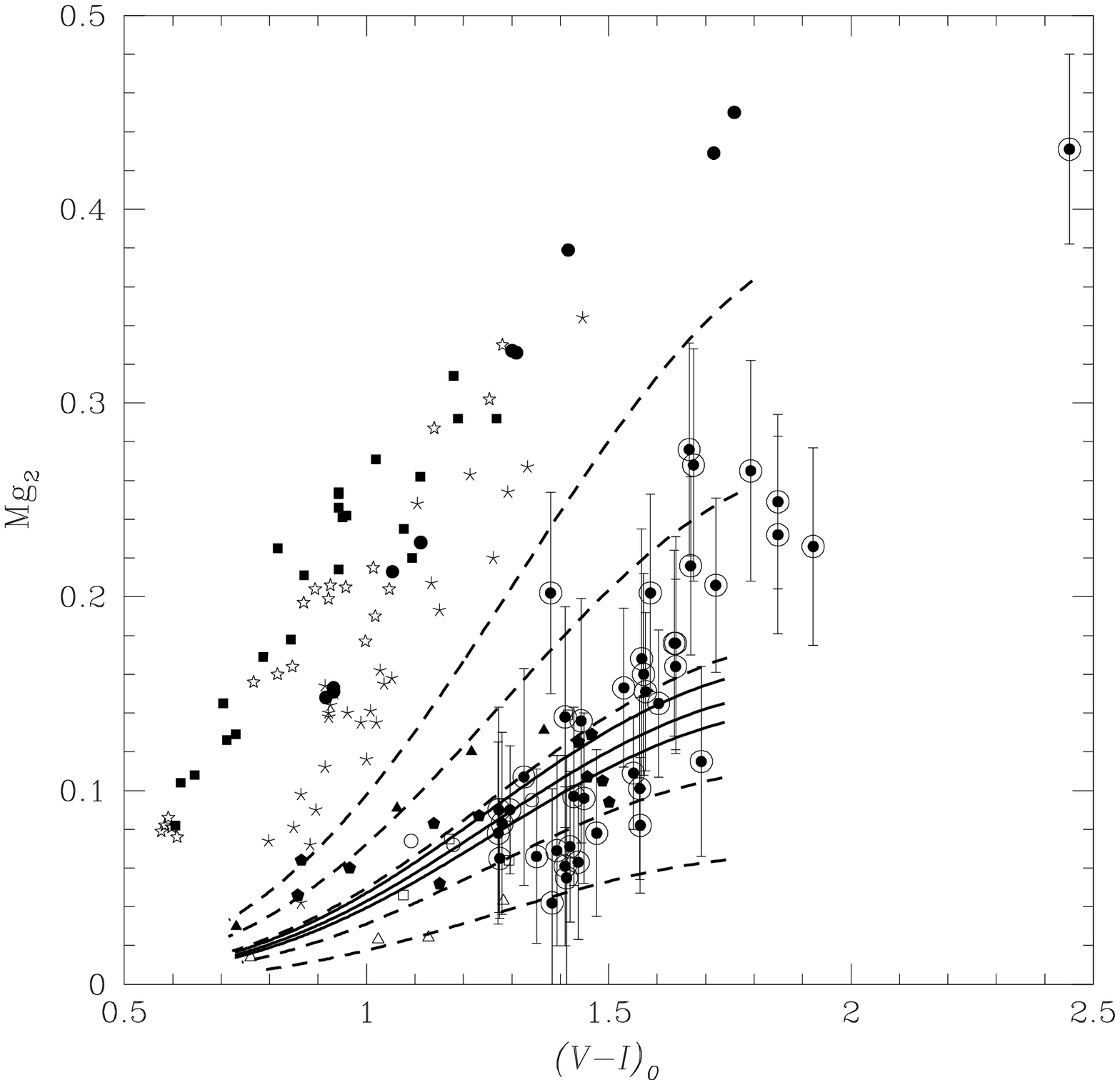}

\figcaption[and2b.10.ps]{Same as Figure 8, except for the Mg$_2$ index.
\label{fig10}}

\clearpage

\plotone{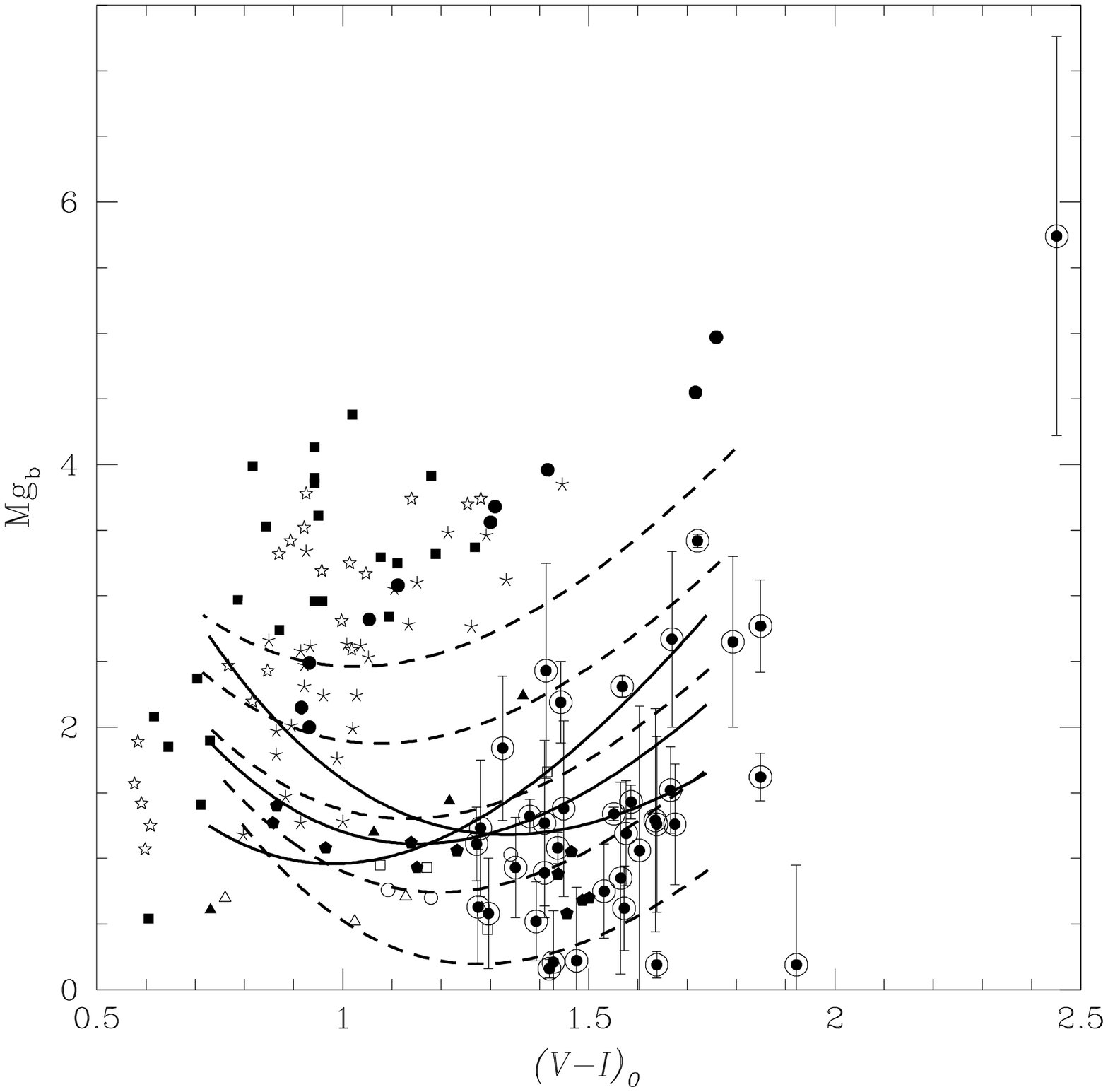}

\figcaption[and2b.11.ps]{Same as Figure 8, except for the Mg$b$ index.
\label{fig11}}

\clearpage

\plotone{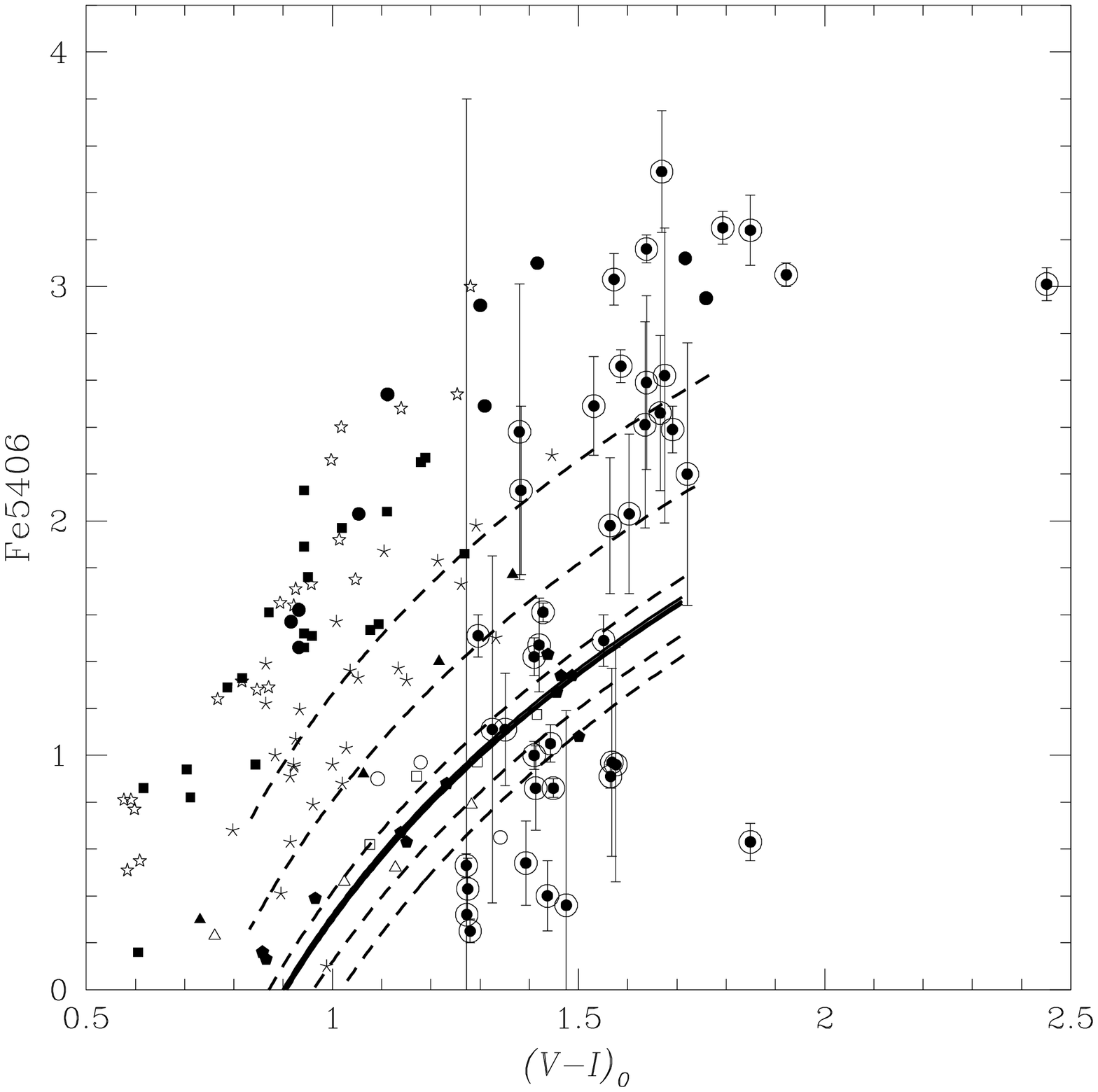}

\figcaption[and2b.12.ps]{Same as Figure 8, except for the Fe5406 index.
\label{fig12}}

\clearpage

\plotone{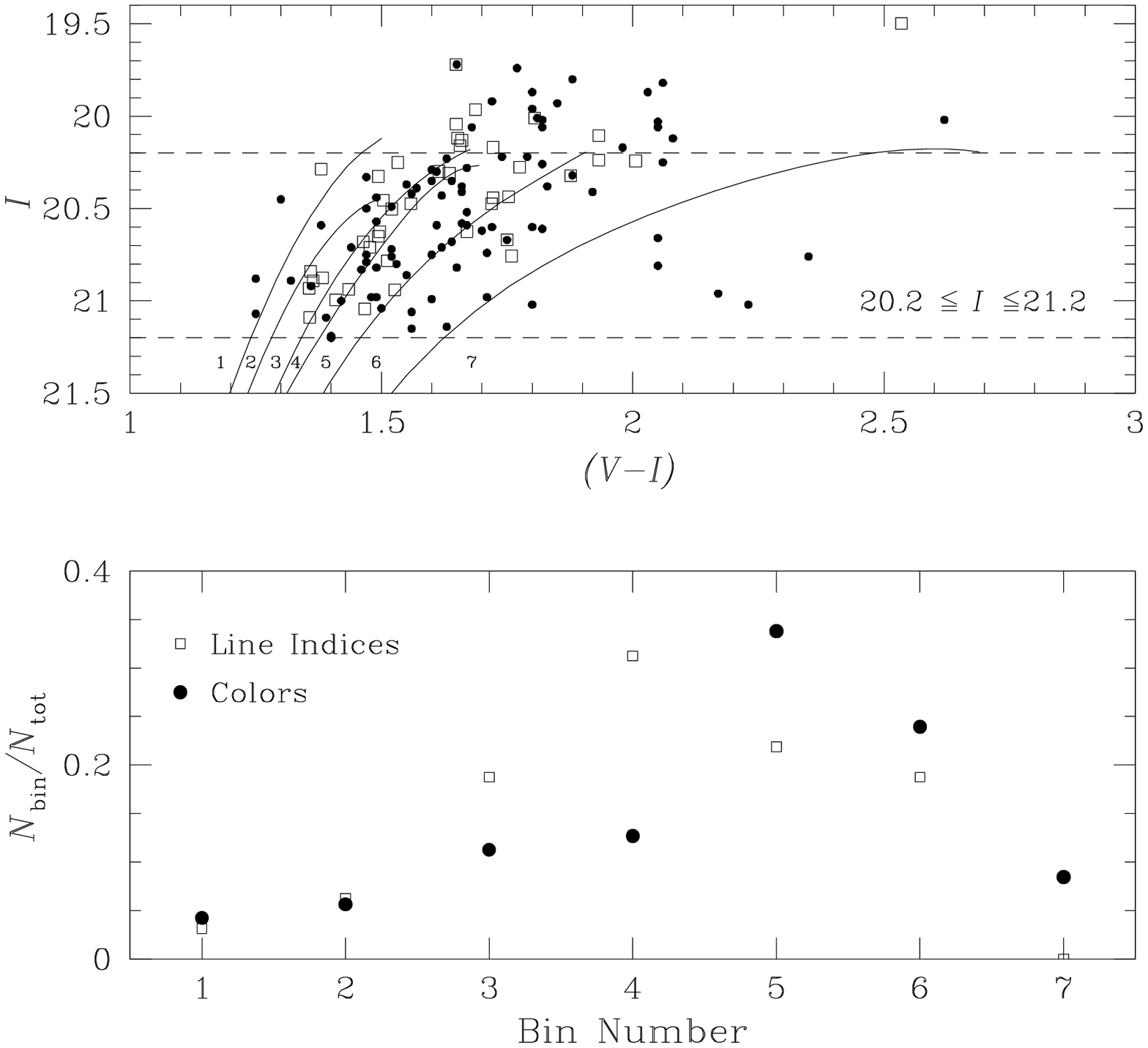}

\figcaption[and2b.13.ps]{(Upper Panel) Distribution in the color-magnitude plane of 
Andromeda II members having
measured line indices (squares) and $20.2 \le I \le 21.2$. Unresolved objects 
within this magnitude interval and located within one arcminute of the galaxy's 
center are shown as circles. The globular cluster fiducial giant branches from 
Figure 3 and 4 are indicated by the curves.
(Lower Panel) Fraction of stars in each metallicity bin (see text).
Squares refer to the stars having measured line 
indices; circles refer the sample of objects having $(V-I)$ colors only.
\label{fig13}}

\clearpage

\plotone{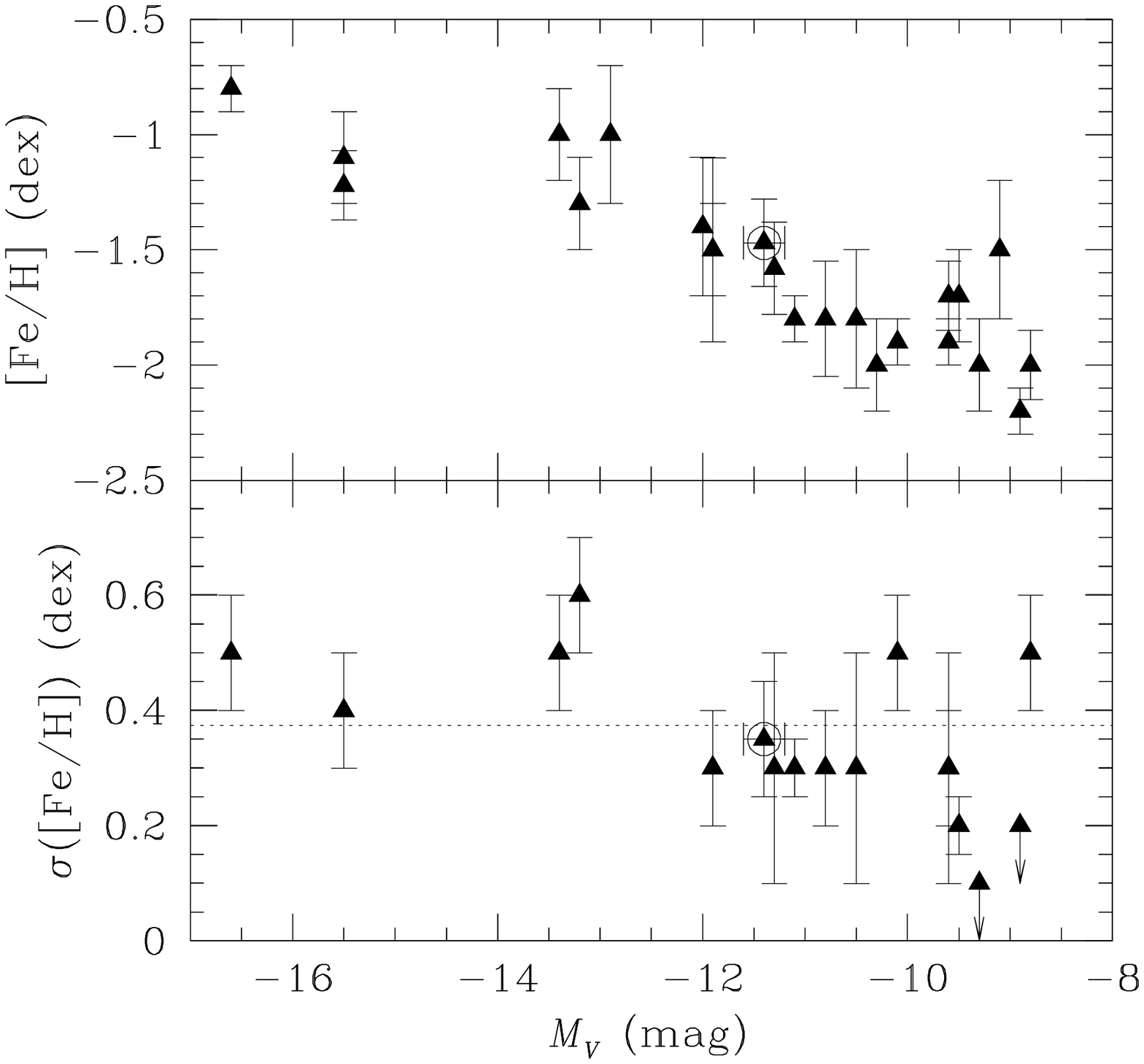}

\figcaption[and2b.14.ps]{(Upper Panel) Mean stellar metallicity plotted against absolute
magnitude for Local Group dE, dSph and dIrr/dSph galaxies. The data are from 
Mateo (1998), Armandroff et al. (1998; 1999), Grebel \& Guhathakurta (1999) and Caldwell
(1999). Andromeda II is indicated by the circled point. 
(Lower Panel) Intrinsic dispersion in metallicity plotted against absolute magnitude 
for Local Group dE, dSph and dIrr/dSph galaxies. The data are from Mateo (1998). The 
circled point indicates the location of Andromeda II. The dotted line indicates the 
mean dispersion of $\sigma({\rm {[Fe/H]}}) = 0.37\pm0.03$ dex (mean error).
\label{fig14}}

 
\end{document}